\documentclass[final,3p,times,twocolumn,fixfloat]{elsarticle}

\usepackage{graphicx}
\usepackage{amssymb} 
\usepackage{amsmath} 
\usepackage{times}
\usepackage{tabularx}

\usepackage{txfonts}
\usepackage{lineno}

\journal{Physics Letters B}

 \usepackage{ifpdf}
 \ifpdf
 \usepackage[pdftex]{hyperref}
 \else
 \usepackage[hypertex]{hyperref}
 \fi

 \hypersetup{
   pdftitle={},%
   pdfauthor={},%
   pdfsubject={},%
   pdfkeywords={},%
   pdfstartview={},%
   bookmarksopen=true, breaklinks=true, debug=true, %
   colorlinks=true, linkcolor=red, citecolor=blue, urlcolor=blue
 }

\usepackage{float} 
\usepackage[caption=false]{subfig} 

\usepackage{amsmath}
\usepackage{epsfig}
\usepackage{float}
\usepackage{verbatim} 
\usepackage{color} 

\newcommand{\ep}{\varepsilon}

\newcommand{\eqs}[1]{\begin{equation} \begin{split} #1\end{split} \end{equation} }

\newcommand{\ie}{{\it i.e.}}
\newcommand{\eg}{{\it e.g.}}
\newcommand{\etal}{{\it et al.}}
\newcommand{\Q}{{\cal Q}}

\newcommand{\ce}[1]{Eq.~(\ref{#1})}

\newcommand{\cf}[1]{{Fig.~\ref{#1}}}

\newcommand{\ct}[1]{{Table~(\ref{#1})}}

\newcommand{\nn}{\nonumber}
\usepackage{multirow} 

\begin{document} 
\begin{frontmatter}

\title{${J/\psi}$-Pair Production at Large Momenta: \\ Indications for Double Parton Scatterings and
Large ${\alpha_s^5}$ Contributions}

\author[IPNO]{Jean-Philippe Lansberg}
\author[PKU,CERN]{Hua-Sheng Shao}
\address[IPNO]{IPNO, Universit\'e Paris-Sud, CNRS/IN2P3, F-91406, Orsay, France}
\address[PKU]{Department of Physics and State Key Laboratory of Nuclear Physics and Technology,
Peking University, Beijing 100871, China}
\address[CERN]{PH Department, TH Unit, CERN, CH-1211, Geneva 23, Switzerland}


\begin{abstract}
\small
We demonstrate that the recent studies of  $J/\psi$-pair production by CMS at the LHC and by D0 at the Tevatron 
reveal the presence of different production mechanisms in different kinematical regions. 
We find out that next-to-leading-order single parton scattering contributions at $\alpha_s^5$ 
dominate the yield at large transverse momenta of the pair. Our analysis further
emphasises the importance of double parton scatterings --which are expected to dominate the yield 
at large $J/\psi$-rapidity differences-- at large invariant masses of 
the pair in the CMS acceptance, and thereby solve a large discrepancy between the theory and the CMS data. 
In addition, we provide the first exact --gauge-invariant and infrared-safe-- evaluation
of a class of leading-$P_T$ ($P_T^{-4}$) next-to-next-to-leading-order contributions at $\alpha_s^6$, 
which can be relevant in the region of large values of $P_{T\rm min}=\min(P_{T1},P_{T2})$.
Finally, we derive simple relations for the feed-down fractions from the production of an
excited charmonium state with a $J/\psi$ in the case of the dominance of the double parton scatterings, 
which significantly deviate from those for single parton scatterings. Such relations can be used
to discriminate these extreme scenari, either DPS or SPS dominance.
\end{abstract}


\end{frontmatter}

\section{Introduction} 
\vspace*{-0.25cm}

Heavy-quarkonium production has attracted 
considerable interest in the high-energy physics community since the $J/\psi$ discovery, exactly forty years
ago. It indeed probes the strong interaction at the interplay of its perturbative and non-perturbative 
regimes~\cite{review}. It can also help to understand a new dynamics of hadron collision
where multiple (hard) parton scatterings (MPS) take place.
MPS are normally very rare since already a single (hard) parton scattering (SPS) is  rare as 
compared to soft scatterings. Owing to the high parton flux at high energies, 
MPS should be likelier at the LHC, starting with two short-distance interactions from a single 
hadron-hadron collision, usually referred to as  double parton scattering (DPS). These have been 
searched in $4$-jets~\cite{Akesson:1986iv,Alitti:1991rd,Abe:1993rv}, $\gamma+3$-jets~\cite{Abe:1997xk,Abazov:2009gc}, 
$W+2$-jets~\cite{Aad:2013bjm,Chatrchyan:2013xxa}, $J/\psi+W$~\cite{Aad:2014rua}, $J/\psi+Z$~\cite{Aad:2014kba}, 
$4$-charm~\cite{Aaij:2012dz}, $J/\psi+$charm~\cite{Aaij:2012dz} and $J/\psi+J/\psi$~\cite{Abazov:2014qba} final states. 

Along these lines, $J/\psi$-pair hadroproduction is of great interest. First, it provides 
an original tool to study quarkonium production in conventional SPSs.
Most of the earlier theoretical studies are based on 
SPSs~\cite{Kartvelishvili:1984ur,Humpert:1983yj,Vogt:1995tf,Li:2009ug,Qiao:2009kg,Ko:2010xy,Berezhnoy:2011xy,Li:2013csa,Lansberg:2013qka,Sun:2014gca};
some using the colour-singlet model (CSM)~\cite{CSM_hadron}, others Nonrelativistic QCD (NRQCD)~\cite{Bodwin:1994jh}. 
Moreover, it is widely claimed that  DPSs~\cite{Kom:2011bd,Baranov:2011ch,Berezhnoy:2012xq,Baranov:2012re} 
could indeed be a significant source of $J/\psi$ pairs at the LHC in proton-proton collisions and in 
proton-nucleus/nucleus-nucleus collisions~\cite{d'Enterria:2013ck,d'Enterria:2014dva}. Generally, it 
remains a poorly understood process. Its measurement with both $J/\psi$ decaying into a muon pair is a 
clean signal, accessible to most experiments, which is complementary to the DPS studies based 
on open charm mesons and hadronic jets. With respect to the latter, it allows one to investigate
the physics of DPS at lower scales and lower $x$ where different mechanisms may be at work 
(see \eg~\cite{Blok:2013bpa}).

The first observation of $J/\psi$-pair events dates back to that of the 
CERN-NA3 Collaboration~\cite{Badier:1982ae,Badier:1985ri}. Recently, the LHCb~\cite{Aaij:2011yc}, 
CMS~\cite{Khachatryan:2014iia} and D0~\cite{Abazov:2014qba} collaborations reported their measurements 
at the LHC and the Tevatron. 
\begin{figure*}[hbt!]
\centering
\subfloat[]{\includegraphics[width=.295\columnwidth,draft=false]{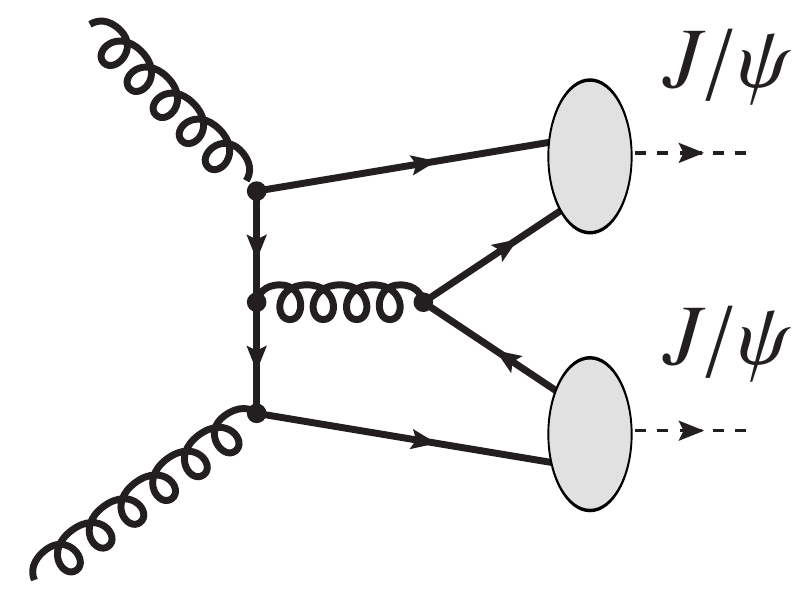}\label{diagram-a}}
\subfloat[]{\includegraphics[width=.295\columnwidth,draft=false]{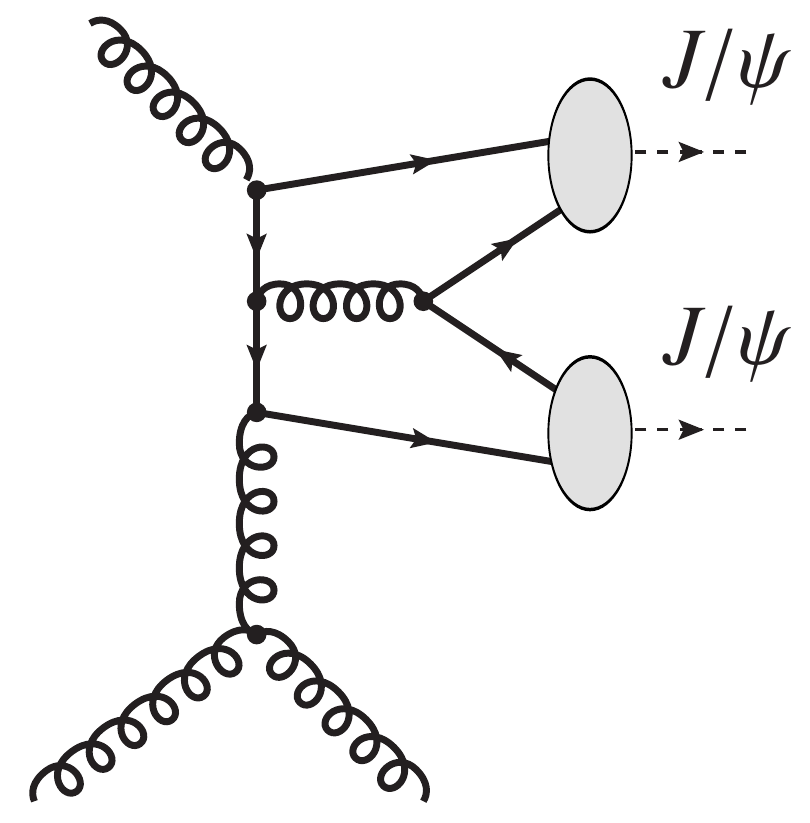}\label{diagram-c}}
\subfloat[]{\includegraphics[width=.295\columnwidth,draft=false]{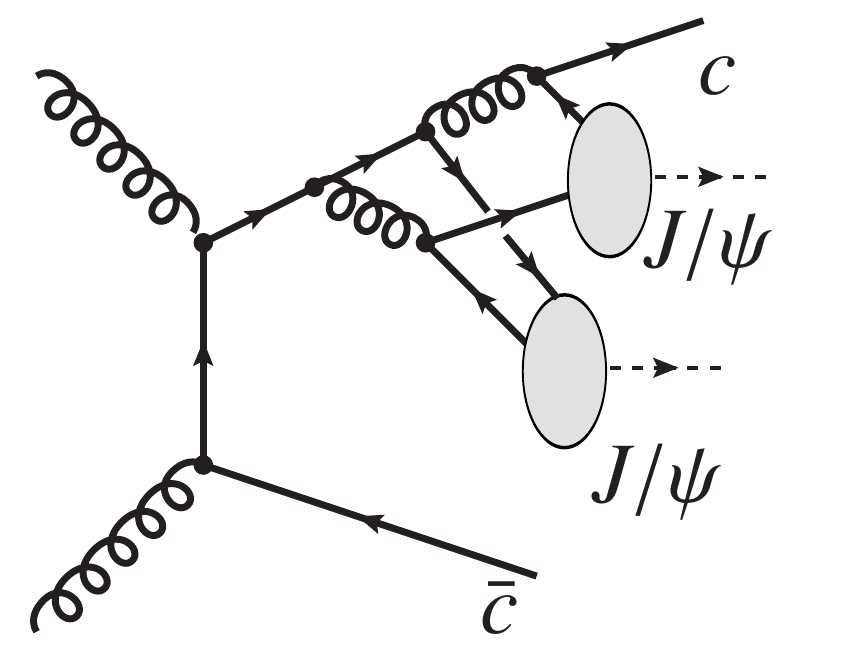}\label{diagram-e}}
\subfloat[]{\includegraphics[width=.295\columnwidth,draft=false]{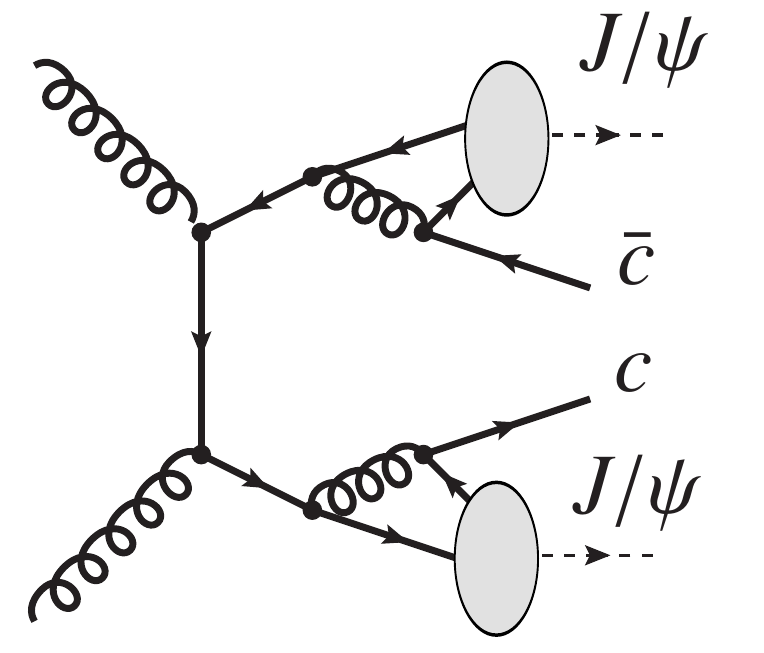}\label{diagram-f}}
\subfloat[]{\includegraphics[width=.295\columnwidth,draft=false]{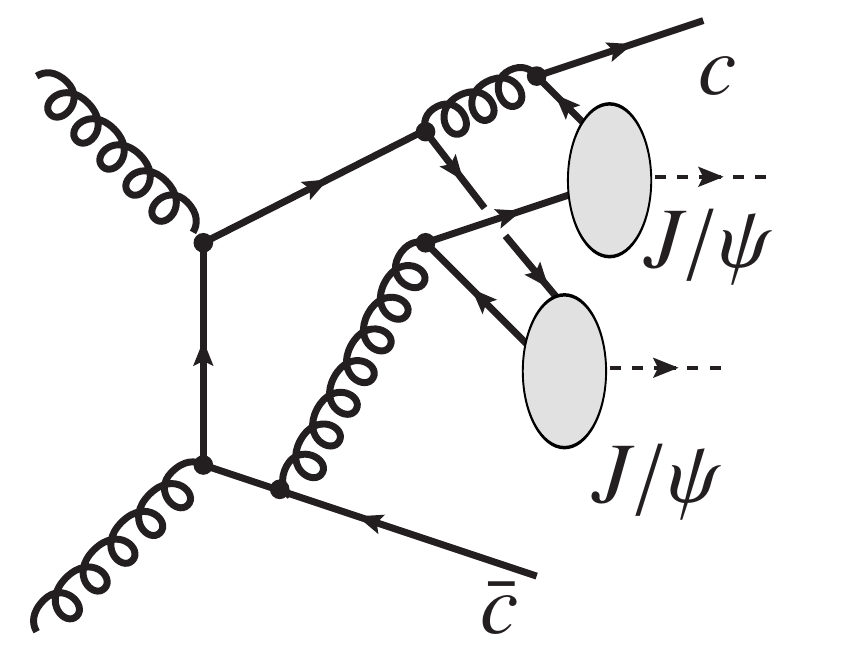}\label{diagram-f2}}
\subfloat[]{\includegraphics[width=.295\columnwidth,draft=false]{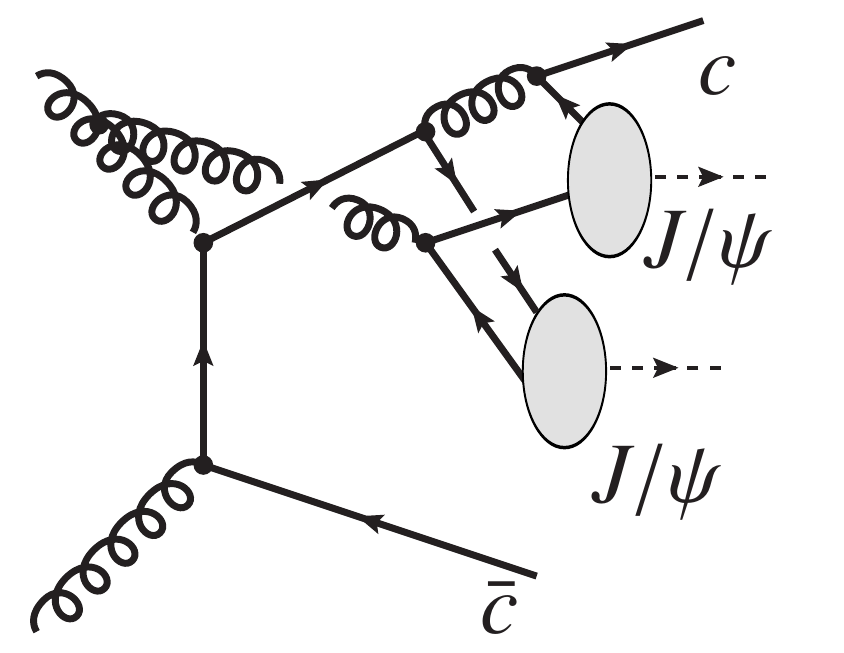}\label{diagram-f3}}
\subfloat[]{\includegraphics[width=.295\columnwidth,draft=false]{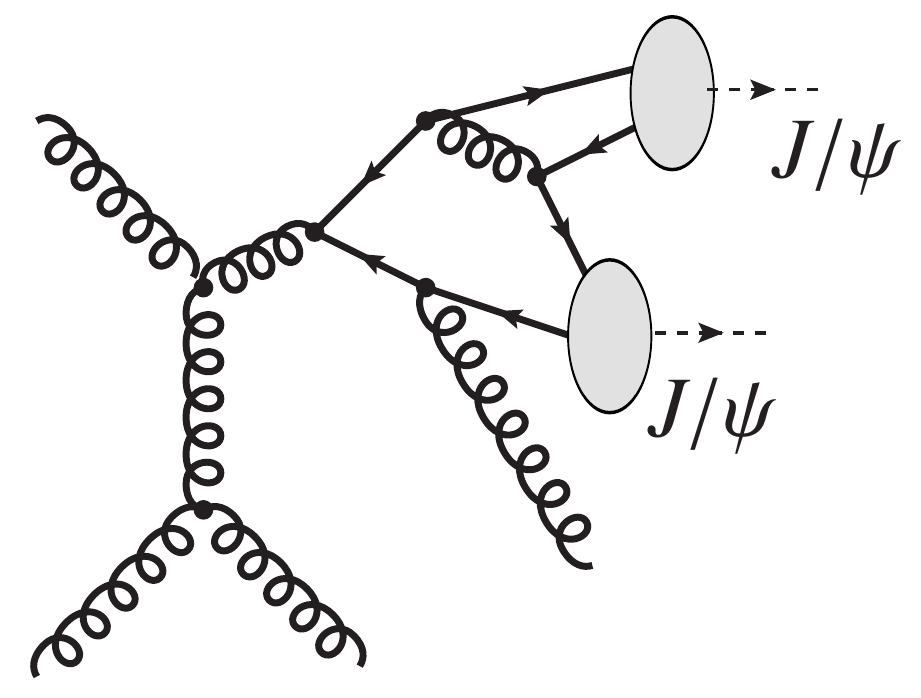}\label{diagram-g}}
\caption{Representative diagrams for the hadroproduction of $J/\psi+J/\psi$ via SPSs at $\mathcal{O}(\alpha_s^4)$ (a), at $\mathcal{O}(\alpha_s^5)$ (b) and 
 $\mathcal{O}(\alpha_s^6)$ (c-g). }
\label{diagrams} \vspace*{-0.5cm}
\end{figure*}
In contrast to Kom~\etal~\cite{Kom:2011bd}, we recently pointed out~\cite{Lansberg:2013qka} that 
 no definite conclusion on the presence of DPSs in LHCb data~\cite{Aaij:2011yc} should be drawn 
given the very large theoretical uncertainties on the SPS predictions. However, the recent 
D0~\cite{Abazov:2014qba} study could provide the very first separation  of the DPSs from SPSs 
and a measurement of $\sigma^{\rm DPS}_{\psi\psi}$ and $\sigma^{\rm SPS}_{\psi\psi}$ by using the 
yield dependence on the (pseudo)rapidity difference between the $J/\psi$ pair, as it 
was first proposed in~\cite{Kom:2011bd}. Although such a separation relies on a good modelling 
of the DPS and SPS rapidity-difference spectra, this can reasonably be considered as the first 
observation of a DPS signal in quarkonium-pair production, even if SPSs also contribute in a 
significant fraction of the D0 acceptance.  Two fundamental remaining questions are whether such 
DPS contributions are also of importance elsewhere than at large rapidity difference, $\Delta y$,
and whether they agree with theory.
In addition, the recent CMS analysis, up to large $J/\psi$-pair transverse momenta ($P_T^{\psi \psi}$) 
brought to light a new striking puzzle. As pointed out in~\cite{Sun:2014gca},
the $P_T^{\psi \psi}$ and invariant mass, $M_{\psi \psi}$, spectra measured by CMS~\cite{Khachatryan:2014iia} 
severely overshoot the SPS contributions --even at next-to-leading order (NLO), \ie~$\alpha_s^5$ . 

In this Letter, we first show that the SPS yield extracted by D0 can only be reproduced thanks to the 
additional $\alpha_s^5$ or feed-down contributions from $J/\psi+\psi'$.
Then, along the lines of~\cite{Kom:2011bd}, we model the DPS spectra based on 3 parametrisations of existing 
single $J/\psi$ data and extract -- accounting for the predicted SPS yield up to  $\alpha_s^5$-- 
from a fit to the CMS results~\cite{Khachatryan:2014iia}
the effective cross section $\sigma_{\rm eff}$ which characterises the effective spatial area of 
the parton-parton interactions. Our fit result is then found to be well compatible with the DPS D0 results~\cite{Abazov:2014qba}
which means that we {\it de facto} provide a solution to the aforementioned puzzle, with a coherent description 
of CMS and D0 results.

In addition, we provide an original test of the DPS vs SPS dominance based on the yields involving
excited states. Such a test can be used to validate our explanation of the CMS puzzle. 
Finally, we evaluate the first piece of the next-to-next-to-leading-order  contributions 
from $gg\to J/\psi J/\psi c \bar c$ (denoted $\psi c \bar c \psi$) which is gauge invariant
and infrared finite. Although it was expected to be enhanced at large 
$P^\psi_{T\rm min}=\min(P^\psi_{T1},P^\psi_{T2})$, we find it is dominant 
only when the yields are out of reach for current experiments and we conclude that an evaluation up to
$\alpha_s^5$ accuracy is probably sufficient at present time.

\vspace*{-0.25cm}
\section{Theoretical frameworks}
\label{sec:theory}
\subsection{SPS Contribution to $J/\psi+J/\psi$ production}

In this section, we outline the computation of the SPS contribution in the CSM~\cite{CSM_hadron} or 
equally NRQCD at LO in $v^2$. The amplitude to produce of a pair of $S$-wave 
quarkonia denoted ${\Q_1}$ and  ${\Q_2}$, of given momenta $P_{1,2}$ and of polarisation $\lambda_{1,2}$ 
accompanied by other partons --inclusive production--, noted $k$, is then given by the product 
of (i) the amplitude to create the corresponding double heavy-quark pair, in the specific kinematical 
configuration where the relative momenta of these heavy quarks ($p_{1,2}$) in each pairs is zero, (ii) 
two spin projectors $N(\lambda_{1,2}| s_{1,3},s_{2,4})$ and (iii) $R_{1,2}(0)$, 
the radial wave functions at the origin in the configuration space for both quarkonia. 
At the partonic level, the SPS amplitude thus reads: 
\eqs{\label{CSMderiv3}
&{\cal M}(ab \to {\Q_1}^{\lambda_1}(P_1)+{\Q_2}^{\lambda_2}(P_2)+k)=\\
&\sum_{s_1,s_2,c_1,c_2}\sum_{s_3,s_4,c_3,c_4}\!\!\!\!\frac{N(\lambda_1| s_1,s_2)N(\lambda_2| s_3,s_4)}{ \sqrt{m_{\Q_1}m_{\Q_2}}} \frac{\delta^{c_1c_2}\delta^{c_3c_4}}{3}\frac{R_1(0)R_2(0)}{4 \pi}\\
&\times {\cal M}(ab \to Q^{s_1}_{c_1} \bar Q^{s_2}_{c_2}(\mathbf{p_1}=\mathbf{0}) + Q^{s_3}_{c_3} \bar Q^{s_4}_{c_4}(\mathbf{p_2}=\mathbf{0}) + k),\nonumber}
where one defines from the heavy-quark momenta, $q_{1,2,3,4}$, 
$P_{1,2}=q_{1,3}+q_{2,4}$, $p_{1,2}=(q_{1,3}-q_{2,4})/2$, and where $s_{1,3}$,$s_{2,4}$ are the 
heavy-quark spin components and $\delta^{c_ic_j}/\sqrt{3}$  is the colour projector. $N(\lambda| s_i,s_j)$ 
is the spin projector, which has a simple expression in the non-relativistic limit: 
$\frac{1}{2 \sqrt{2} m_{Q} } \bar{v} (\frac{\mathbf{P}}{2},s_j) \Gamma_{S} u (\frac{\mathbf{P}}{2},s_i) \,\, $ 
where $m_Q$ is the heavy-quark mass and $\Gamma_S$ is $\ep^{\lambda}_{\mu}\gamma^{\mu}$ when $S=1$ 
(\eg~ $J/\psi$, $\psi'$). Such a partonic amplitude is then squared, summed over the
colour and spin of external partons and convoluted with the partonic densites (PDFs) in the allowed
kinematical phase space. 

In the case of $gg\to J/\psi J/\psi+c\bar c$, there exist more than 
2000 graphs (see \cf{diagrams} (c-f)) 
which are non-trivially zero even after the topologies with a single gluon connected to an individual 
heavy-quark lines are removed. In order to generate the amplitude for this process, and the other ones 
considered in this study, we use {\sc HELAC-Onia}, described in~\cite{Shao:2012iz}, which generates
the amplitude based on \ce{CSMderiv3} using recursion methods \cite{Kanaki:2000ey}. It can also deal with 
$P$-wave production which involves the derivative of the amplitude in the relative momentum of the heavy-quark
forming the quarkonia.

{\sc HELAC-Onia} also performs the helicity-amplitude calculations and the convolution with the PDF 
as well as the final-state variable integration. At LO, $J/\psi$-pair production at colliders is 
from $gg\to J/\psi+J/\psi$ at $\alpha_s^4$ (see \eg~\cf{diagram-a}). At $\alpha_s^5$, one needs to 
consider real-emission contribution (see \eg~\cf{diagram-c}) as well as loop corrections.
In~\cite{Sun:2014gca}, a full NLO computation  showed that for\footnote{Unless specified otherwise,  $P^\psi_T$
is the $P_T$ of one $J/\psi$ randomly chosen among both and the $P_T$ cuts discussed here apply to both.}  
$P^\psi_T > 2$ GeV, it is sufficient 
to rely on a NLO$^\star$~\cite{Lansberg:2013qka} evaluation where only the LO and NLO real-emission 
contributions are accounted for; the latter being regulated by an infrared cut-off. 
This is easily explained by the $P_T^2$ suppression of the loop contributions. 
For the $\psi c \bar c \psi$ contribution which we compute, there is no need to apply any infrared cut-off. 
Since the LO kinematics is that of a $2\to 2$ process (\cf{diagram-a}), it generates a trivial 
$P_T^{\psi \psi}$ dependence ($\delta(P_T^{\psi \psi})$) in the collinear factorisation with
conventional PDFs. We have therefore accounted for the $k_T$'s of the initial partons with a Gaussian 
distribution with $\langle k_T \rangle=2$~GeV as in~\cite{Lansberg:2013qka} in order to obtain
fairer comparisons of the $P_T^{\psi \psi}$ spectra. Technical details about the implementation
 can be found in~\cite{Shao:2015vga}. 

As regards the PDFs, we use the set CTEQ6L1 for  LO (${\cal O}(\alpha_s^4)$) calculations and 
CTEQ6M~\cite{Pumplin:2002vw} for NLO$^{\star}$ (${\cal O}(\alpha_s^5)$)  and the ${\cal O}(\alpha_s^6)$ 
$\psi c \bar c \psi$ calculations. The SPS uncertainties are obtained by a common variation of $m_c$
and $\mu_F=\mu_R$ as $((1.4\hbox{ GeV},0.5 \times m_T^{\psi\psi});(1.5\hbox{ GeV}, m_T^{\psi\psi});(1.6\hbox{ GeV},2 \times m_T^{\psi\psi}))$
with $m_T^{\psi\psi}=\big((\sum m_i)^2+(P_T^{\psi})^2\big)^{1/2}$ 
with $\sum m_i =4 m_c$, but for $\psi c \bar c \psi$ where  $\sum m_i =6 m_c$.

\subsection{DPS  Contribution to $J/\psi+J/\psi$ production}
\label{subsec:DPS}

Quarkonium-pair (${\cal Q}_1+{\cal Q}_2$) production from DPSs is usually assumed to come 
from 2 independent SPSs which create each a single quarkonia. One therefore assumes
\begin{eqnarray}
\sigma^{\rm DPS}_{{\cal Q}_1{\cal Q}_2}=\frac{m}{2}\frac{\sigma_{{\cal Q}_1}\sigma_{{\cal Q}_2}}{\sigma_{\rm eff}},
\label{eq:DPS-fact}
\end{eqnarray}
where $m=1$ for identical final-state particles and $m=2$ otherwise, $\sigma_{\cal Q}$ is the cross section 
for {\it single} ${\cal Q}$ production. $\sigma_{\rm eff}$ is expected to account for the effective
size of the parton-parton interaction and should thus be universal --that is process-independent--  
as well as $\sqrt{s}$-independent if the factorisation holds as in~\ce{eq:DPS-fact}. 
Yet, there does not exist proofs of such a factorisation. Factorisation-breaking effects
have been discussed in a number of recent studies (see \eg~\cite{Blok:2013bpa,Kasemets:2012pr,Diehl:2014vaa}).  
As of today, data is needed to test it case by case. Finally $\sigma_{\rm eff}$ cannot be determined 
from first principles or from perturbative methods.

Anticipating the discussion of the D0 results in the next sections, which extracted a
smaller $\sigma_{\rm eff}$ from $J/\psi$-pair production than usually found in previous studies
involving jet observables, we note that $\sigma_{\rm eff}$ could very well depend
on the flavour of the initial partons (see \eg~\cite{Blok:2013bpa}). In the present case, these 
are gluons only, whereas for high-$P_T$ jets with $W$, light-quark initiated processes give a 
significant contribution. In addition, the process considered here occurs at a rather low 
momentum scale. Finally, we stress that if the 1v2 contributions\footnote{Such contributions, also
sometimes denoted $1 \otimes 2$, arise from 2 parton-parton scatterings 
where 2 partons from one proton come from a single parton. In a sense, these only initially involve 3 partons.} 
discussed in~\cite{Gaunt:2014rua} matter, this would result
in larger DPS contributions, thus in a smaller extracted $\sigma_{\rm eff}$.

Following the common practice, DPS yields are simply computed from the corresponding measured 
single-$J/\psi$ yields using a Monte Carlo code with as input a parametrisation of $\sigma_{\psi}$, 
see \eg~\cite{Kom:2011bd}. Let us stress here that if one uses a parametrisation of 
prompt, \ie\ excluding $b$-decay $J/\psi$, or direct quarkonium data, one would predict DPS yields for 
prompt or direct quarkonium pairs. Given the level of understanding of quarkonium-production mechanisms, 
using theoretical models to compute $\sigma_{\psi}$ entering \ce{eq:DPS-fact}  would only inflate 
the theoretical uncertainties. As for now, the objective of DPS studies is to quantify their impact 
and to verify the factorisation hypothesis in a given kinematical domain. In the present study, 
the objective is for instance to address the apparent discrepancy between the predicted SPS yield and the CMS data.

\begin{figure}[h!]
\begin{center}
\includegraphics[width=.875\columnwidth,draft=false]{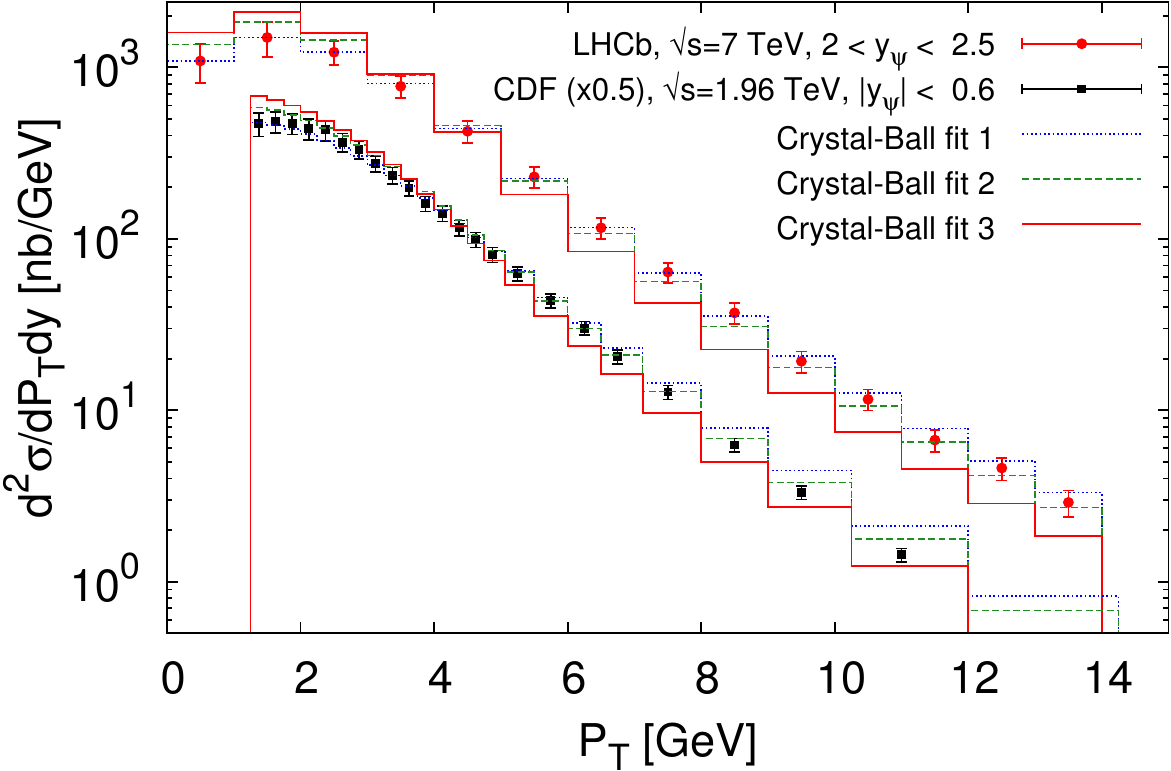}
\caption{Illustrative comparison between  3 fits of $\sigma_\psi$  and LHCb~\cite{Aaij:2011jh} and CDF~\cite{Acosta:2004yw} 
data for prompt $J/\psi$ production.}\label{fig:comparison-fit}\vspace*{-0.5cm}
\end{center}
\end{figure}

We thus use the setup proposed in~\cite{Kom:2011bd} for the single $J/\psi$ cross section, $\sigma_\psi$, 
(Eqs.(2-4) of~\cite{Kom:2011bd} slightly improved since we used 3 $\sigma_\psi$ fits 
in order to assess the systematic uncertainties attached to the parametrisation of $\sigma_\psi$.
Details regarding these fits are given in the \ref{appendix-fits}. As an illustrative
comparison, the fits to LHC and Tevatron data are shown on~\cf{fig:comparison-fit}. We
stress that the data used for the 3 fits are for prompt $J/\psi$.
The corresponding short-distance matrix element has been added to a specific branch of 
{\sc HELAC-Onia}~\cite{Shao:2012iz} with as inputs the fit parameters. This branch 
is separate from that used to compute the SPS contributions (for technical details, the reader is referred to~\cite{Shao:2015vga}). 

We stress that the purpose
of using an event generator such as {\sc HELAC-Onia} for DPS computations is to perform the spin-entangled decay of the $J/\psi$'s 
under different polarisation hypotheses to
apply the fiducial cuts (on the muons) of a given analysis if the muon acceptance was not corrected, 
as for the D0 analysis~\cite{Abazov:2014qba}. A simple combination of $2 \sigma_{\psi}$ would not allow for this. 
In the D0 case, a variation of $\lambda_\theta$ within $-0.45 <\lambda_\theta<0.45$, which represents
 a reasonable envelope of the existing experimental measurements in similar conditions (see \eg~\cite{Abulencia:2007us}),
induces a systematical 20$\%$ uncertainty. This can be compared to the 25 \% systematical uncertainty on the corrected muon 
acceptance quoted by CMS~\cite{Khachatryan:2014iia} due to the unknown $J/\psi$ polarisation.

As in~\cite{Kom:2011bd}, we use the MSTW 2008 NLO PDF set~\cite{Martin:2009iq} 
and the factorisation scale $\mu_F=m_T^\psi=(m_{\psi}^2+(P_T^{\psi})^2)^{1/2}$. 
The fit uncertainties attached to our DPS evaluation are discussed in the next sections.
Finally, let us mention that we have studied one factorisation-breaking effect which is the 
possible correlation between two partons from a single proton as encoded in the double PDF 
(dPDF) of \cite{Gaunt:2009re}.
We did not find any relevant difference in the region considered in this study.

\vspace*{-0.25cm}
\section{Feed-down relations for the DPS and SPS yields}

\subsection{Feed-down fractions under DPS dominance}
\label{subsec:FD_DPS}

If one sticks to a simplistic --although widely used-- view of the DPS production mechanisms 
as the one presented above, it is possible to derive general relations between the 
feed-down fractions of the DPS yields for double and single $J/\psi$ production. These can be used 
to evaluate the feed-down impact, but also, by returning the argument, to test a possible 
DPS-dominance hypothesis by directly measuring pair productions involving the excited states.

Just as we define the fractions, $F^{\rm direct}_\psi$, $F^{\chi_c}_\psi$ and $F^{\psi'}_\psi$,  
of single  $J/\psi$ produced directly, from $\chi_c$ decay or from $\psi'$ decay, 
 one can  define various feed-down fractions  for $J/\psi+J/\psi$. However, 
one should keep in mind that it would  probably be experimentally very challenging
to measure (and subtract) the yield of $\chi_c+\chi_c$ or even $\chi_c+\psi'$. We therefore limit ourselves to define 
$F^{\chi_c}_{\psi\psi}$ (resp. $F^{\psi'}_{\psi\psi}$)
as the fraction of  $J/\psi+J/\psi$ events from the feed-down of {\it at least} a $\chi_c$ (or resp.  a $\psi'$) decay. 
In other words, $F^{\chi_c}_{\psi\psi}$ is the fraction of events including a prompt $J/\psi$ (direct or from 
$\chi_c$ and $\psi'$ feed-down) plus a $J/\psi$ identified as from a $\chi_c$.  Although it is probably 
very difficult to measure it, we also define $F^{\rm direct}_{\psi\psi}$ as being the pure  direct 
component, excluding all the possible feed-downs, which can be easier to predict theoretically.

\begin{table*}[hbt!] 
  \begin{center} \renewcommand{\arraystretch}{1.8}\setlength\tabcolsep{5pt}\small
\begin{tabularx}{17.5cm}{cp{5cm}|cccccc} 
\hline  \hline      & Energy and quarkonium cuts   &$\sigma_{\rm exp.}$ &$\sigma^{\rm SPS, prompt}_{\rm LO}$ &
$\sigma^{\rm SPS, prompt}_{\rm NLO^{(\star)}}$ & $\sigma^{\rm DPS, prompt}$ & $\chi^2$ \\
\hline\hline 
LHCb
 &  $\sqrt{s}=7$~TeV, $P^{\psi_{1,2}}_T < 10$~GeV, \hspace*{0.82cm} $2<y_\psi < 5$~\cite{Aaij:2011yc} &$18 \pm 5.3$ pb  & $41^{+51}_{-24}$  pb  & $46^{+58}_{-27}$ &   $31^{+11}_{-6.3}(^{+24}_{-15})$ pb &$0.5 - 1.2$\\\hline 
\multirow{2}{*}{D0} & $\sqrt{s}=1.96$~TeV, $P^{\psi_{1,2}}_T > 4$~GeV, & SPS: $70 \pm 23$ fb  & $53^{+57}_{-27}$ fb 
& $170^{+340}_{-110}$ fb &  -- & --\\ 
                    & $|\eta_{\psi}|<2.0$ ~\cite{Abazov:2014qba} (+ $\mu$ cuts in caption)& DPS: $59 \pm 23$ fb           &        --           & --  & 
$44^{+16}_{-9.1}(^{+7.5}_{-5.1})$ fb&    $0.06 - 0.5$                            \\ \hline 
CMS  & $\sqrt{s}=7$~TeV, $P_T^{\psi_{1,2}}> 6.5 \rightarrow 4.5$~GeV depending on $|y_{\psi_{1,2}}| \in [0,2.2]$ (see the caption)~\cite{Khachatryan:2014iia}&$5.25\pm 0.52$  pb & $0.35^{+0.26}_{-0.17}$  pb  & $1.5^{+2.2}_{-0.87}$  pb & $0.69^{+0.24}_{-0.14}(^{+0.039}_{-0.027})$ pb & $1.09 - 1.14$ \\\hline 
ATLAS & $\sqrt{s}=7$ TeV,  $P_T^{\psi_{1,2}}>5$ GeV and $|y_{\psi_{1,2}}|<2.1$ (+ $\mu$ cuts in the caption)~\cite{private-ATLAS}& -- & $6.4^{+4.3}_{-2.6}$ fb & $36^{+49}_{-20}$ fb &  $19^{+6.8}_{-4.0}(^{+2.2}_{-1.6})$ fb& N/A\\ \hline\hline 
\end{tabularx}
\caption{Comparison for $\sigma(pp(\bar{p})\to J/\psi+J/\psi+X) \times {\cal B}^2(J/\psi \to \mu\mu)$ between the LHCb, CMS and D0 data and our predictions  in the relevant kinematical regions (+ that of the forthcoming ATLAS analysis).
The theory predictions are: the SPS prompt yields at LO and NLO$^{\star}$ [For LHCb, the evaluation is a complete NLO~\cite{Sun:2014gca}], 
the DPS prompt yields with $\sigma_{\rm eff}$ fitted to the CMS differential distributions (see section~\ref{subsec:CMS-data}) and the $\chi^2$ between the sum of DPS+SPS (resp. DPS) yield and CMS and LHCb (resp. D0 DPS) data. For the DPS yields, 
the first uncertainty is from $\sigma_{\rm eff}$ (see~\ct{fitDPS}) and the second in parenthesis 
is a systematical certainty from the 3 fits (alike the variation of the central value of $\sigma_{\rm eff}$ 
in~\ct{fitDPS}).
The range of the $\chi^2$  also comes from the 3 fits.
[The additional uncorrected $\mu$ cuts are: for D0, $P_T^{\mu}> 2$~GeV when $|\eta_{\mu}|<1.35$ and total momenta $|p^{\mu}|>4$~GeV when $1.35<|\eta_{\mu}|<2.0$; for ATLAS: 
$P_T^{\mu}>2.5$~GeV and  $|\eta_{\mu}|<2.3$ and at least one $J/\psi$ with two muons with $P_T^{\mu}>4$ GeV. For CMS, the detailed cuts are $P_T^{\psi}> 6.5$~GeV if $|y_{\psi}|<1.2$; $P_T^{\psi}> 6.5 \rightarrow 4.5$~GeV 
if $1.2<|y_{\psi}|<1.43$;$P_T^{\psi}> 4.5$~GeV if $1.43<|y_{\psi}|<2.2$ where in $1.2<|y_{\psi}|<1.43$, 
the $P_T^{\psi}$ cutoff scales linearly with $|y_{\psi}|$].
 }\vspace*{-0.5cm}
\label{xsections}
\end{center}
\end{table*}

Assuming \ce{eq:DPS-fact} holds for all charmonia, one gets\footnote{The derivation of \ce{eq:FD_fraction_DPS}
for $\chi_c$ follows from the decomposition of the different sources of a prompt $J/\psi$ + a $J/\psi$ from a $\chi_c$. Namely, one has~: direct $J/\psi+\chi_c$, $\chi_c+\chi_c$ and $\psi'+\chi_c$. Their cross section with the relevant branchings can then be decomposed in terms of single quarkonium cross sections using \ce{eq:DPS-fact} taking care of not double counting $\chi_c+\chi_c$ ($m=1$). Their sum divided by the cross section for a pair of prompt $J/\psi$ decomposed likewise then reads as \ce{eq:FD_fraction_DPS} using the standard definitions of $F^{\chi_c}_\psi$, $F^{\psi'}_\psi$ and $F^{\rm direct}_\psi$.}
\eqs{\label{eq:FD_fraction_DPS}
F^{\chi_c}_{\psi\psi}&= F^{\chi_c}_\psi \times \big(F^{\chi_c}_\psi + 2 F^{\rm direct}_\psi + 2 F^{\psi'}_\psi\big),\\
F^{\psi'}_{\psi\psi}&= F^{\psi'}_\psi \times \big(F^{\psi'}_\psi + 2 F^{\rm direct}_\psi + 2 F^{\chi_c}_\psi\big),\\
F^{\rm direct}_{\psi\psi}&= (F^{\rm direct}_\psi)^2.}
In order to obtain numbers,  let us recall that the world data tell us that 
$F^{\rm direct}_\psi$, $F^{\chi_c}_\psi$ and $F^{\psi'}_\psi$ are close to 60\%, 30\% and 10\%. 
We then obtain $F^{\chi_c}_{\psi\psi} \simeq 50 \%$, $F^{\psi'}_{\psi\psi} \simeq 20\%$ 
and $F^{\rm direct}_{\psi\psi} \simeq 35 \%$. Although $F^{\chi_c}_{\psi\psi}$ and $F^{\psi'}_{\psi\psi}$ 
are experimentally accessible via $\sigma((\chi_c \to J/\psi)+J/\psi)$ and $\sigma((\psi' \to J/\psi)+J/\psi)$, 
they are not sufficient to determine the pure direct yield since 
$F^{\rm direct}_{\psi\psi} \neq 1 - F^{\chi_c}_{\psi\psi} - F^{\psi'}_{\psi\psi}$. Its extraction would require
the measurement of $\sigma(\chi_c+\psi')$.

\subsection{Feed-down fractions under SPS dominance}

On the contrary, one expects a larger feed-down from $\psi'$ if SPSs dominate.
In the CSM or NRQCD at LO in $v^2$, the hard part for $\psi'+J/\psi$ and $J/\psi+J/\psi$ is identical; only $|R(0)|^2$ differ.
Taking $|R_{\psi'}(0)|^2=0.53~{\rm GeV}^3$~\cite{Eichten:1995ch}, whereas $|R_{J/\psi}(0)|^2=0.81~{\rm GeV}^3$,
and ${\cal B}(\psi'\to J/\psi)=55 \%$~\cite{Nakamura:2010zzi} as well as accounting for 
a factor $2$ from the final-state symmetry, the ratio of $F^{\psi'}_{\psi\psi}/F^{\rm direct}_{\psi\psi}$ --defined as in section \ref{subsec:FD_DPS}-- is expected
to be as large as $0.53/0.81 \times 0.55 \times 2 + (0.53/0.81 \times 0.55)^2\simeq 0.85$. 
It may even be a bit larger since we neglected $\sigma(\chi_c+\psi')$ in this evaluation of $F^{\psi'}_{\psi\psi}$. 
The latter approximation is justified, since we checked that neither $\sigma(\chi_c+J/\psi)$ nor 
$\sigma(\chi_c+\psi')$ are significant under SPS dominance. In the CSM they are absent at $\alpha_s^4$. The colour-octet 
(CO) contributions for the production of these pairs are small because
of the small size of the CO non-perturbative parameters (also called LDMEs)~\cite{Shao:2014fca} 
and the absence of any kinematical enhancement.
In the remaining our this work, we will thus consider that $\sigma_{\rm SPS}^{\rm prompt}=
1.85 \times \sigma_{\rm SPS}^{\rm direct}$. In turn, we also have  $F^{\psi'}_{\psi\psi}\simeq 0.85/(1+0.85) \simeq 46 \%$
at any order in $\alpha_s$.

\subsection{DPS vs. SPS}

To summarise, in the SPS case,  $F^{\psi'}_{\psi\psi}$ can be as large as $46 \%$ whereas $F^{\chi_c}_{\psi\psi}$ 
is expected to be small. In the DPS case, $F^{\psi'}_{\psi\psi}$ is half as small, around $20\%$,
and $F^{\chi_c}_{\psi\psi}$ large, around $50\%$.
This clearly means that the relative measurements of charmonium-pair 
production of different states can serve as a clear test to pin down DPS or SPS dominance since
they correspond to rather opposite predictions. This can reliably be done provided that
the single-charmonium yields are known in the same kinematical region. We stress that, for
such a test, we do not need to know the value of $\sigma_{\rm eff}$ which does not appear in \ce{eq:FD_fraction_DPS}.

\vspace*{-0.25cm}
\section{Data-theory comparisons}

\subsection{The early LHCb data at low transverse momenta}
 Let us first look at the LHCb data. We claimed in a recent work~\cite{Lansberg:2013qka} that there was no compelling reason 
to call for significant DPS contributions in order to describe the $J/\psi$-pair measurement by LHCb at 7 TeV
in the forward rapidity region ($2 < y_\psi <4.5$). In particular, there is absolutely no difficulty
to reproduce the measured yield with the SPS contributions alone, see the first line of \ct{xsections}. In fact, 
the LO~\cite{Lansberg:2013qka} and NLO~\cite{Sun:2014gca} prompt SPS values  
even tends to be above the LHCb one, leaving room for a possible DPS yield {\it only} when the uncertainties
are accounted for. We stress that this measurement was performed 
without any lower $P_T$ cuts and that, in this case, the LO and NLO SPS predictions 
are in very good agreement, showing a good convergence of the perturbative series.
We will comment later on the DPS predictions.

\subsection{The D0 data up to large $\Delta y$ and the observation of DPS contributions}
We now discuss the recent D0 measurement~\cite{Abazov:2014qba}. 
Thanks to a wide (pseudo)rapidity coverage of about 4 units ($|\eta|\leq 2$), the D0 detector made 
the first extraction of the DPS contributions to  $J/\psi$-pair production possible.
As neatly discussed in~\cite{Kom:2011bd}, the yield as a function of the rapidity difference $\Delta y$ 
between both $J/\psi$'s should be a good observable to distinguish DPS and SPS events. The DPS
 events have a broader distribution in $\Delta y$ than the SPS ones. For the latter, large values of 
$\Delta y$ imply large momentum transfers, thus highly off-shell particles, and are strongly suppressed. It is not 
the case for DPSs where the rapidity of both $J/\psi$ is independent. 
Large rapidity differences are only suppressed because the individual yields are suppressed for increasing rapidities.

By fitting the  $\Delta y$  
distribution of their data, D0 managed~\cite{Abazov:2014qba} to separate out the DPS and the SPS yields, \ie\
 $\sigma^{\rm DPS}_{\psi\psi}$ and $\sigma^{\rm SPS}_{\psi\psi}$.
They found that about half of the (prompt) yield was from SPSs, 
the remaining half from DPSs, about 60 fb (see \ct{xsections}).

In~\cite{Lansberg:2013qka}, we discussed 
the relevance of taking into account $P_T$-enhanced 
topologies (e.g. \cf{diagram-c}) at NLO and performed, for the first time, a partial NLO evaluation, dubbed
as NLO$^\star$.
Indeed, the real $\mathcal{O}(\alpha_s^5)$ emissions should be dominant in the intermediate 
and high $P_T$ regimes. This was recently confirmed by a full NLO evaluation~\cite{Sun:2014gca}: 
the NLO$^\star$ $P_T^{\psi}$ spectrum indeed accurately coincides with the NLO one for $P_T^{\psi}>2 m_c$. 
It is important to note that the NLO$^{(\star)}$ yield is almost one order of magnitude larger than LO one 
when $P_T^{\psi} = 5~{\rm GeV}$ and that one must use
a NLO (or NLO$^\star$) evaluation when dealing with data sets with a $P_T$ cut as it is the case here for all
but LHCb data. 

Since both $J/\psi$ should have their $P_T> 4$ GeV, it is interesting to look at 
the impact of the $\alpha_s^5$ corrections (NLO$^\star$). Whereas the LO SPS yield 
is a bit below the SPS D0 yield, the NLO$^\star$ yield, which is about 3 times larger, is above. 
Both agree with the data within the large theoretical uncertainties --mainly from $m_c$. 
As we shall see later, the need for $\alpha_s^5$ corrections is far more obvious in the CMS acceptance
with a higher $P_T$ cut.

Injecting their measured $\sigma_\psi$ and $\sigma^{\rm DPS}_{\psi\psi}$ in \ce{eq:DPS-fact},
 D0 has found $\sigma_{\rm eff} \simeq 5.0 \pm 2.75~{\rm mb}$. This value is 3 times lower
than those extracted with jet observables, which means that the DPS yield seems to be 
3 times higher than what could have been naively expected -- or at least twice higher accounting
from their uncertainty. Yet, as next section will show, a low value of $\sigma_{\rm eff}$
allows one to solve the CMS puzzle.

\begin{figure}[t!]
\begin{center}
\subfloat{\includegraphics[width=\columnwidth,draft=false]{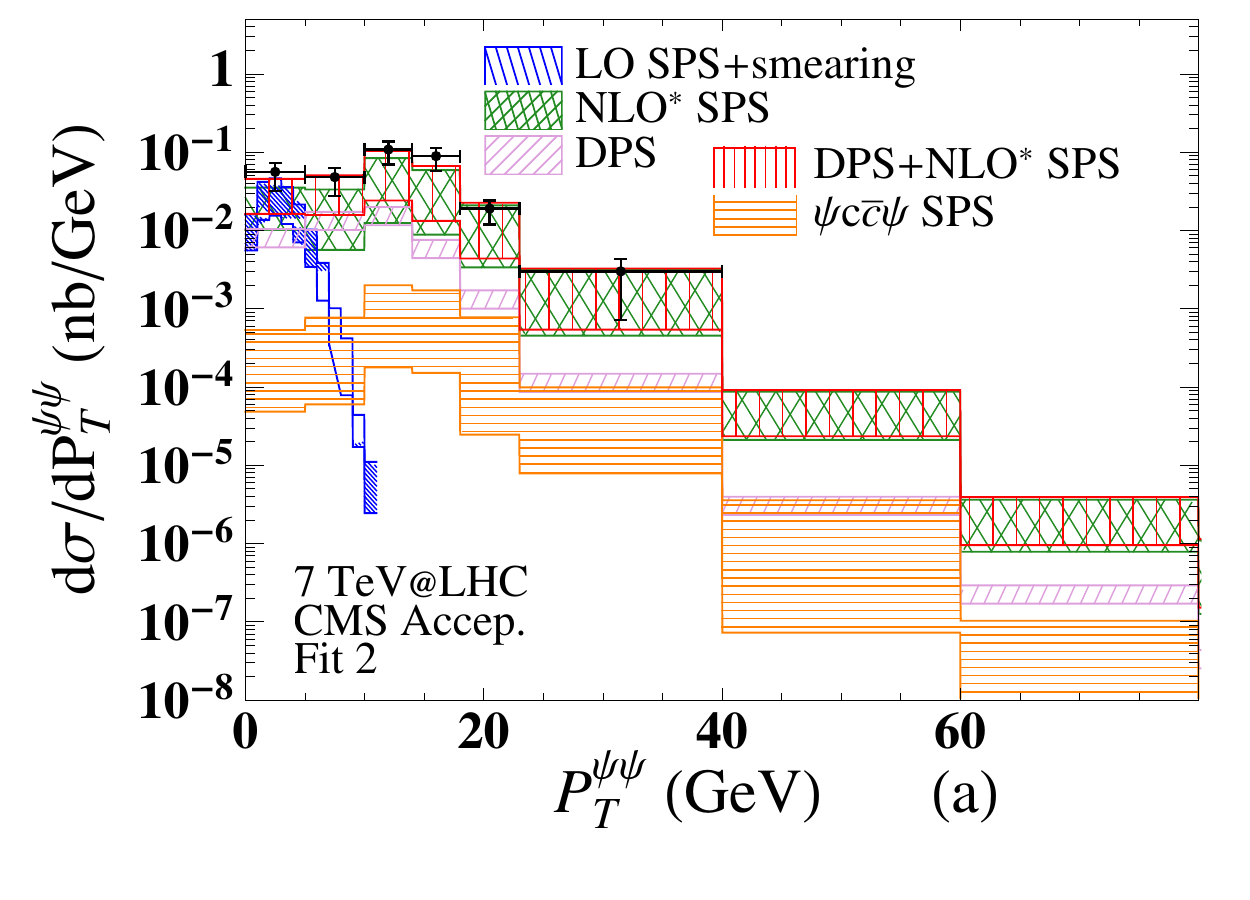}\label{fig:dsigCMSa}}\vspace*{-0.25cm}\\
\subfloat{\includegraphics[width=\columnwidth,draft=false]{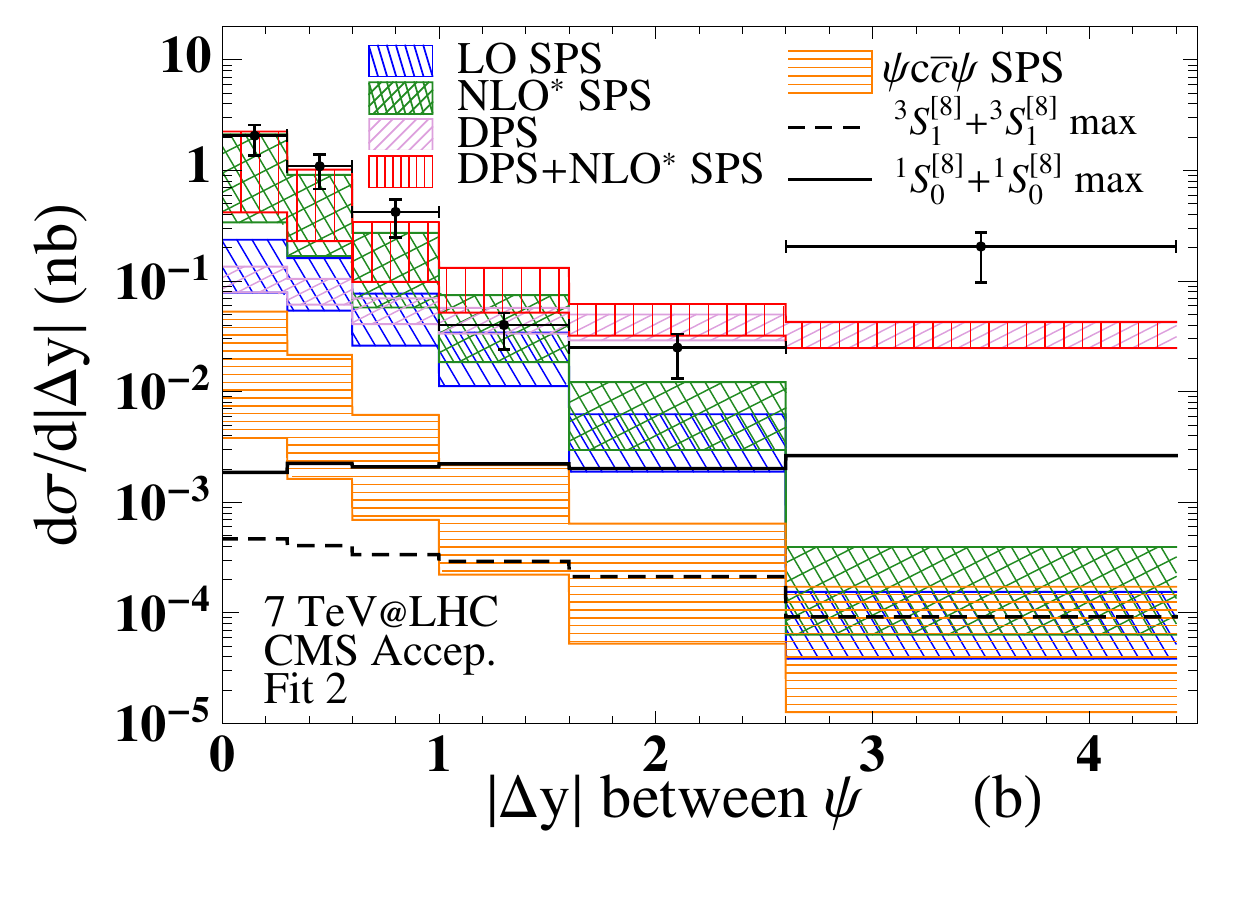}\label{fig:dsigCMSb}}\vspace*{-0.25cm}\\
\subfloat{\includegraphics[width=\columnwidth,draft=false]{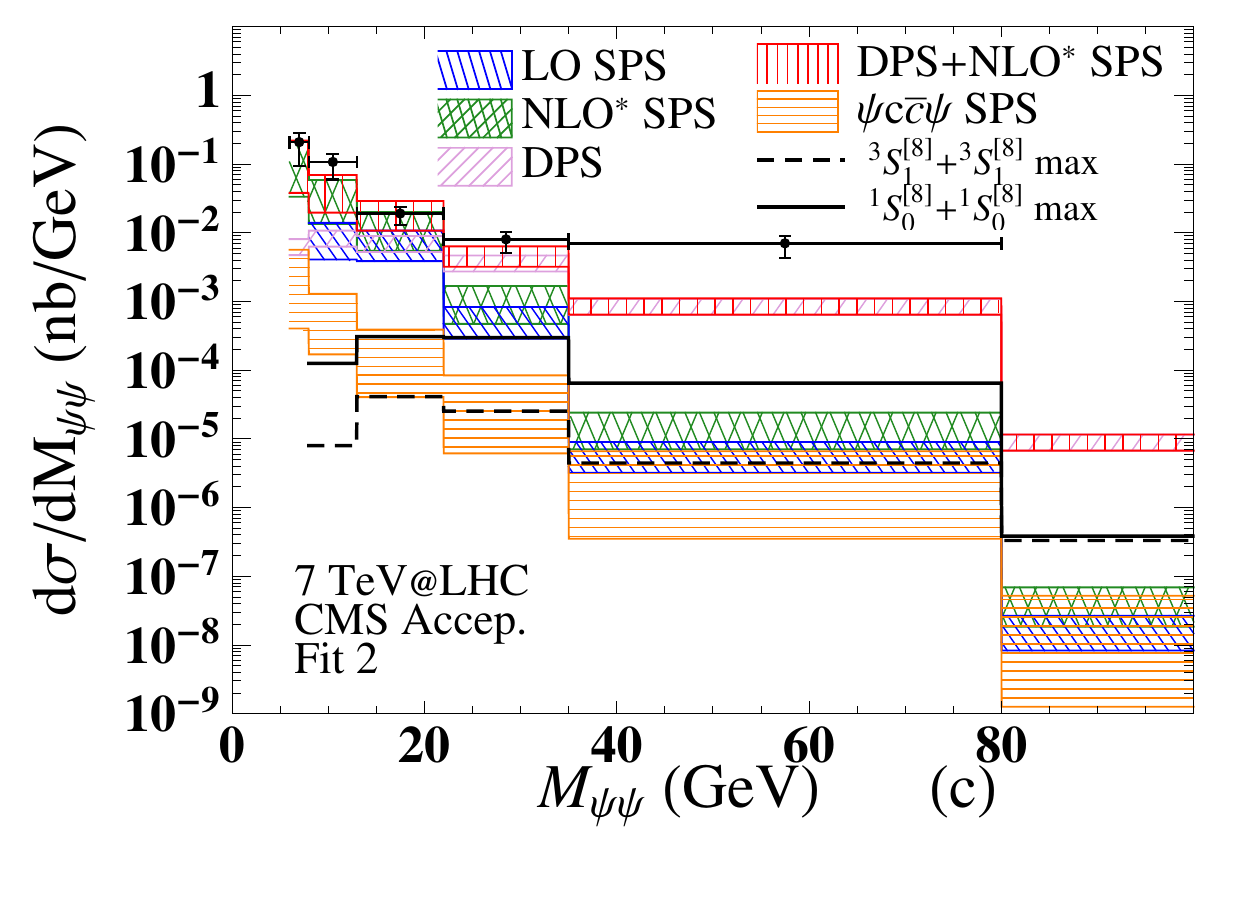}\label{fig:dsigCMSc}}\vspace*{-0.25cm}\\

\caption{Comparison of different theoretical contributions with the CMS measurement: (a)  pair transverse momentum; (b) absolute-rapidity difference ; (c) pair invariant mass.}
\label{fig:CompareCMS}
\end{center}\vspace*{-1cm}
\end{figure}

\subsection{The CMS data at large momenta} 
\label{subsec:CMS-data}

LO and NLO$^\star$ SPS cross sections for  prompt $J/\psi$ pair production in the 
CMS acceptance~\cite{Khachatryan:2014iia} are given in \ct{xsections}.
As expected because of the higher $P_T$ cut, one observes a larger  NLO$^\star$/LO ratio than in the D0
acceptance. Yet the NLO$^\star$ SPS yield is significantly below the CMS data~\cite{Khachatryan:2014iia}
and hint at the presence of another source of $J/\psi$ pairs.
As we whall see, the discrepancy is much more evident when one looks at 
differential distributions.

Indeed, besides this integrated yield, CMS measured differential distributions~\cite{Khachatryan:2014iia}
which further indicate the importance of both NLO SPSs and DPSs but in different regions.
Overall CMS released in addition 17 data points of differential cross sections as a function of 
$P_T^{\psi \psi}$, $|\Delta y|$ and $M_{\psi \psi}$. To quantify the impact
of the DPS contributions, we have used these experimental data to fit $\sigma_{\rm eff}$ via \ce{eq:DPS-fact}
using the 3 fits of $\sigma_\psi$ discussed in section~\ref{subsec:DPS} and subtracting our theoretical 
evaluations of the SPS NLO$^\star$ yield acounting for their uncertainties (green band in the plot of \cf{fig:CompareCMS}).

\ct{fitDPS} summarises the fit result: the $\chi^2_{\rm d.o.f.}$ and the values of $\sigma_{\rm eff}$ along with 
their uncertainites coming from (a) the CMS experimental uncertainties\footnote{including
the 25 \% systematical uncertainty due to the unknown $J/\psi$ polarisation as discussed above.}  and (b) the theoretical uncertainties on the SPS yield.
We have also given the $\chi^2_{\rm d.o.f.}$ when no DPS contribution is considered. We note that the goodness of the 3 
fits is similar.  The dispersion of the central values of $\sigma_{\rm eff}$ thus allows us to assess a systematical uncertainty due to
the paramatrisation of $\sigma_\psi$. \cf{fig:CompareCMS} shows the DPS distributions with the Fit~2. Comparison plots using the Fit~1 \&~3 are
given a supplementary materials. In addition, we note that $\sigma_{\rm eff}$ fitted using the Fit~3 (only Tevatron data) 
is very close to the D0 value. Whereas the $p$-values\footnote{assuming Gaussian uncertainties which is probably not true for 
theory and some systematical experimental uncertainties.} of our DPS fits of $\sigma_{\rm eff}$ are about 2\% (see also below), the one without 
DPS is below 0.03\% (even less without $\alpha_s^5$ contributions).

\begin{table}
\begin{center}  \small 
\begin{tabular}{c|cccccccccc}
       & $\sigma_{\rm eff}$ [mb]& $\chi^2_{\rm d.o.f.}$ & d.o.f. \\
\hline\hline
$\sigma_\psi$  Fit 1 \cite{Kom:2011bd}   & $11  \pm 2.9$  &  1.9 & 16 \\
$\sigma_\psi$ Fit 2                      & $8.2 \pm 2.2$  &  1.8 & 16 \\
$\sigma_\psi$ Fit 3                      & $5.3 \pm 1.4$  &  1.9 & 16 \\
Only LO SPS                             & N/A            &  7.6 & 17 \\
Only NLO$^\star$ SPS                      & N/A           &   2.6 & 17 \\
\end{tabular}\vspace*{-0.5cm}
\end{center}
\caption{Result of the fit of the DPS yield via $\sigma_{\rm eff}$ on the 18 CMS values.\label{fitDPS}}\vspace*{-0.5cm}
\end{table}

As regards the $J/\psi$-pair $P_T$, $P_T^{\psi \psi}$, distribution, \cf{fig:dsigCMSa}
clearly emphasises the importance of $\alpha_s^5$ QCD corrections to the SPS yield. 
Phenomenologically accounting for the initial parton $k_T$ is not sufficient to reproduce the data
as the smeared LO curve shows. 
At NLO, hard real-emissions tend to generate larger momentum imbalances
and configurations with $P_T^{\psi \psi}\neq 0$. In fact, the real-emission topologies (\cf{diagram-c}) tend to produce,
at large $P_T$, two {\it near} $J/\psi$ --as opposed to back-to-back-- with a large $P_T^{\psi \psi}$.
In the case of DPSs, correlations are absent and configurations at low $P_T^{\psi \psi}$ are favoured. There is also no reason
for large $P_T^{\psi \psi}$ configurations to be --relatively-- enhanced. It is thus not surprising that 
the DPS band drops faster than the NLO$^\star$ SPS one at large $P_T^{\psi \psi}$. 
The ``bump'' around $P_T^{\psi \psi}\simeq12~{\rm GeV}$ simply reflects the kinematic cuts
in the CMS acceptance.  Overall, one obtains a good agreement ($\chi^2_{\rm d.o.f.}\simeq 1.1$) with the $P_T^{\psi \psi}$ distribution when DPS and (NLO$^\star$) SPS 
contributions are considered together, but it also confirms the dominance
of (NLO$^\star$) SPS contributions at large  $P_T^{\psi \psi}$.

In addition, CMS analysed the relative-rapidity spectrum, $d\sigma/d|\Delta y|$. Along the lines of the D0 
data discussion, the SPS contribution dominate when $|\Delta y| \to 0 $, 
while the DPS ones are several orders of magnitude larger than the SPS ones at large $|\Delta y|$. 
A comparison with the CMS data is shown in \cf{fig:dsigCMSb}.
Most of the data are consistent with our results, except for the last bin, 
which probably explains the low $p$-value which we obtained with the DPS fit. For this data set, 
$\chi^2_{\rm d.o.f.}\simeq 2.1$ with DPS for the 3 fits and about 2.6 without (only NLO$^\star$). Note
the large NLO$^\star$ SPS uncertainty which de facto reduces the corresponding $\chi^2$.
The CMS acceptance with a rapidity-dependent $P_T$ cut renders $d\sigma/d|\Delta y|$ 
flatter but this effect is apparently not marked enough in our theory curves. 
More data are however needed to confirm the absence of binning effects which could have generated 
a dip in the distribution. As another possible cross-check, we provide 
predictions for the ATLAS, D0, and LHCb acceptances in \ct{xsections} and as supplementary materials.

At $P_T^{\psi \psi}=0$ --where the bulk of the yield lies--, the $J/\psi$-pair invariant mass, $M_{\psi \psi}$, 
is closely related to $|\Delta y|$ and provides similar information (see~\cf{fig:dsigCMSc}). One indeed has
$M_{\psi \psi}=2 m_T^{\psi}\cosh{\frac{\Delta y}{2}}$. 
Large $\Delta y$ --\ie~large relative {\it longitudinal} momenta-- correspond to 
large $M_{\psi \psi}$. [At $\Delta y=3.5$ and $P_T=6$ GeV, $M_{\psi \psi}\simeq 40$ GeV.] Without additional
cuts, the  $M_{\psi \psi}$  and $d\sigma/d|\Delta y|$ spectra of the CMS do reveal the same conclusion: 
the DPS contributions dominate the region of large momentum differences. At small $M_{\psi \psi}$, 
SPS contributions dominate and NLO corrections are large --essentially because CMS data do not cover low $P^\psi_T$.
For this data set, $\chi^2_{\rm d.o.f.}\simeq 3.0$ with DPS for the 3 fits and about 5.3 without (only NLO$^\star$).

\subsection{Back to D0 and LHCb data}

As we have just seen, the inclusion of the DPS contributions with $\sigma_{\rm eff}$ ranging from 5 to 11 mb, depending
on the fit used for $\sigma_\psi$, solves the so-called CMS puzzle found (with SPS contributions only) in~\cite{Sun:2014gca} with a $\chi^2_{\rm d.o.f.}$ 
reduced for all the distributions. Using these fits of $\sigma_{\rm eff}$, we should also reproduce the D0 DPS rate~\cite{Abazov:2014qba}. 
The agreement (see \ct{xsections}) is quite good ($\chi^2<1$) and gives us confidence
that our proposed solution to the CMS puzzle is indeed correct. The comparison with the LHCb result is less instructive
owing to the uncertainties from the data and from the SPS.

\vspace*{-0.25cm}
\section{A first step toward a NNLO evaluation of the SPS contributions}
At high $P_T^{\psi}$,  $\mathcal{O}(\alpha_s^6)$ (NNLO) contributions like  $g g \to c \bar{c}$ 
followed by $c(\to J/\psi +J/\psi+c)$ (\cf{diagram-e}) or twice by $c(\to J/\psi +c)$  (\cf{diagram-f})
and $g g \to g^\star g \to (J/\psi +J/\psi+g) g$ (\cf{diagram-g}) are expected to be enhanced by factors 
of $P_T^{\psi}$ w.r.t NLO~\cite{Lansberg:2013qka,Sun:2014gca}. A two-loop computation is 
however needed to evaluate them, which is beyond the state-of-the-art. 
Yet, such leading-$P_T$ contributions can be evaluated via the fragmentation approximation, as done in~\cite{Sun:2014gca} 
only for topologies like~\cf{diagram-f}, which were expected to dominate at large $M_{\psi \psi}$. 
However, such an approximation has been shown~\cite{Artoisenet:2007xi} to be unjustified 
for the similar process $gg \to J/\psi c \bar{c}$ unless $P_T$ is much larger than $m_\psi$.
As discussed in section~\ref{sec:theory}, the process $gg\to J/\psi J/\psi+c \bar{c}$ is infrared safe and can be computed by itself
using {\sc HELAC-Onia} ``out-of-the-box''~\cite{Shao:2012iz}. 
As just said, it includes
a class of likely dominant NNLO corrections depicted in \cf{diagram-e} and \cf{diagram-f}. 
On the contrary, the contributions from topologies \cf{diagram-g} could be evaluated using {\sc \small HELAC-Onia}
but only with an infrared cutoff as in \cite{Artoisenet:2008fc,Lansberg:2013qka} for the NNLO$^\star$. 
In the present study, we prefer to limit ourselves to $gg\to J/\psi J/\psi+c \bar{c}$ which does not require
any ad-hoc prescription. 

The band labelled $\psi c \bar c \psi$ in \cf{fig:CompareCMS} shows its full contribution, which is computed for the first time.
This partial $\alpha_s^6$ contribution is as large as the NLO$^\star$ ones only at the highest $M_{\psi \psi}$ and $\Delta y$, 
where the DPS ones are anyhow dominant. The case of another variable, the {\it sub-leading} $P_T$, $P_{T\rm min}$, is however particular since
the DPS spectrum is expected to scale as $P_{T\rm min}^{-2\,\times\, n}$, $n$ being the scaling power of the single $J/\psi$ yield ($P_T^{-n}$).
We thus found $J/\psi J/\psi+c \bar{c}$  to be dominant (see  \cf{fig:dsigCMSd} in the \ref{sec:Ptmin-plot})
at very high $P_{T\rm min}$, which  corresponds to back-to-back production as in \cf{diagram-f}. 

Overall, the aforementioned missing fragmentation contributions (\cf{diagram-g}) at $\mathcal{O}(\alpha_s^6)$ 
are expected to be of similar sizes. In general, a full NNLO computation is thus expected to be similar
to one at NLO accuracy, except in kinematical regions which are not currently accessible. 
Corresponding predictions for the forthcoming ATLAS and LHCb analyses as well as the current D0 acceptance
are given as supplementary material for comparison with forthcoming data.

\vspace*{-0.25cm}
\section{Possible impact of colour-octet transitions}

We have also investigated the possible impact of CO channels as discussed at LO in~\cite{Li:2009ug,Qiao:2009kg}.
We found that, because of the double suppression of the CO LDMEs, CO+CO
 channels are nowhere important when $P_T^{\psi}<50~{\rm GeV}$, as we found out~\cite{Lansberg:2013qka}.
We have evaluated the contribution from  $^3S_1^{[8]}+\ ^3S_1^{[8]}$ and $^1S_0^{[8]}+\ ^1S_0^{[8]}$ (see~\cf{fig:CompareCMS}) 
using the 1-$\sigma$ upper value of the  $^3S_1^{[8]}$ (resp. $^1S_0^{[8]}$) LDMEs
of the NLO prompt fit of \cite{Butenschoen:2011yh}, \ie\ 0.00283 GeV$^3$ (resp. 0.0541 GeV$^3$), these
are compatible with the LO direct fit of \cite{Sharma:2012dy} and are the only ones not  dramatically overshooting
the low-$P_T$ single $J/\psi$ data~\cite{Feng:2015cba}.
As we look for an upper value, we disregard the  $^3P_J^{[8]}+\ ^3P_J^{[8]}$ contribution which is negative. A complete CO study is
beyond the scope of this Letter and will be the object a dedicated publication.

In any case, this upper limit of the CO contributions is always smaller than the CS ones except for $|\Delta y| > 2.5$ (last bin in \cf{fig:dsigCMSb}) and $M_{\psi\psi} > 40$~GeV 
(last two bins in \cf{fig:dsigCMSc}). In these regions, these SPS contributions are anyhow extremely suppressed as compared to DPS ones
and the CS SPS can receive significant $\alpha_s^6$ contributions. 
The only distribution where the CO contributions might show up is for the $P_{T \rm min}$ (\cf{fig:dsigCMSd}); it has
a similar size and the same dependence as our partial NNLO evaluation. They are larger than the NLO$^\star$ SPS and DPS yields
only where the cross sections are on the order of $10^{-8}$ nb before accounting for the branchings.  
As regards the mixed CO+CS channels, there is no $P_T^{\psi}$ enhancement to be expected and 
these are simply suppressed by the LDME. 

\vspace*{-0.25cm}
\section{Conclusion}
In this Letter, we have focused on the explanation of the recent observations of (prompt) $J/\psi$-pair 
production made by D0 and CMS. The measurements by CMS~\cite{Khachatryan:2014iia},
which severely overshoot the theory if one solely considers SPS contributions~\cite{Sun:2014gca},
indicate significant DPS contributions, which we find to agree with the magnitude measured by D0~\cite{Abazov:2014qba}. 
For the first time, our study shows that both DPSs and the NLO 
QCD corrections to SPSs are crucial to account for the existing data.

If these experimental results are confirmed, this would provide evidence
for 
\begin{enumerate}[(i)]
\item the dominance of $\alpha_s^4$ (LO) contributions for the total cross section, 
\item the dominance 
of $\alpha_s^5$ (NLO) contributions at mid and large $P_T^{\psi \psi}$, 
\item  the dominance of DPS contributions
at  large $\Delta y$ and at large $M_{\psi \psi}$. 
\end{enumerate}
We have also derived generic formulae predicting feed-down contributions or, equally speaking, charmonium-pair-production rates involving excited states,
 in case DPSs dominate. These can be checked
by measuring $J/\psi+\psi'$ or $J/\psi+\chi_c$ production. Such data can also therefore check a 
possible DPS dominance as found by D$0$ and CMS at large momenta. The relatively small 
value of $\sigma_{\rm eff}$ (see \cf{fig:sigma_eff}) compared to jet-related extractions we obtained to describe 
the CMS data --also compatible with the D0 DPS yield
and their extracted value of $\sigma_{\rm eff}$-- may
be a first hint at the flavour dependence of this effective cross section.

\begin{figure}[h!]
\begin{center}
\includegraphics[width=\columnwidth,draft=false]{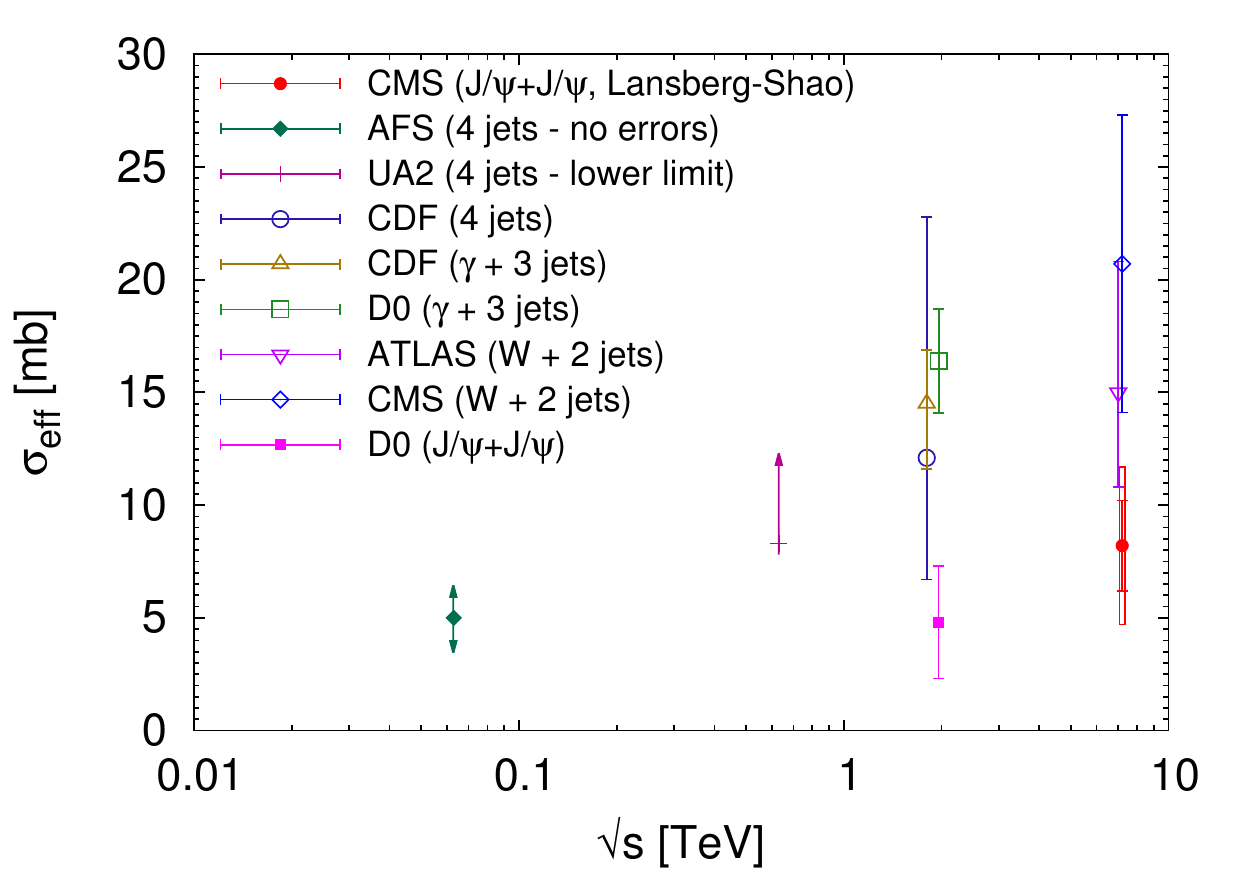}\vspace*{-0.25cm}\\
\caption{Comparison of our CMS fit result of $\sigma_{\rm eff}$ ($8.2 \pm 2.0 \pm 2.9$~mb: first uncertainty 
from the data and the SPS theory uncertainty, the second from the single $J/\psi$ template) with 
other extractions~\protect\cite{Akesson:1986iv,Alitti:1991rd,Abe:1993rv,Abe:1997xk,Abazov:2009gc,Aad:2013bjm,Chatrchyan:2013xxa,Abazov:2014qba}.\label{fig:sigma_eff}}
\label{fig:CompareCMS}
\end{center}
\end{figure}

Finally, we have carried out 
the first exact evaluation of leading-$P_T$ NNLO ($\alpha_s^6$) 
contributions, \ie~$J/\psi$-pair production with a $c\bar c$ pair, which could contribute
at large $P_{T\rm min}$.
On the way, our study of the impact of all the real emissions at $\alpha_s^5$ and some at $\alpha_s^6$  
also demonstrates the absence of a significant colour-octet 
contribution contrary to what was found at LO in~\cite{Li:2009ug,Qiao:2009kg}.

\vspace*{-0.25cm}
\section*{Acknowledgments}
\vspace*{-0.25cm}
\begin{small} 
 We thank D. Bandurin, V. Belyaev, K.-T. Chao, D.~d'Enterria, E.G. Ferreiro, S. Spanier, B. Weinert, Z. Yang and D. Zieminska for fruitful and stimulating 
discussions. This work is 
supported in part by the France-China Particle Physics Laboratory (FCPPL). H.-S. Shao is 
also supported by the National Natural Science Foundation of China (No 11075002, No 11021092) 
and ERC grant 291377 ``LHCtheory: Theoretical predictions and analyses of LHC physics: 
advancing the precision frontier". 
\end{small}



\appendix

\vspace*{-0.5cm}
\section{Single $J/\psi$ fits}
\label{appendix-fits}

Following~\cite{Kom:2011bd}, the  $gg\to \Q+X$ amplitude squared, $|\mathcal{A}_{gg\to\Q+X}|^2$,  is  parametrised by a Crystal Ball function~:\vspace*{-0.25cm}
\eqs{
\left\{
\begin{array}{ll}
\frac{\lambda^2\kappa\hat{s}}{M_{\Q}^2}\exp(-\kappa\frac{P_T^2}{M_{\Q}^2})
& \mbox{when $P_T\leq \langle P_T\rangle$} \\
\frac{\lambda^2\kappa\hat{s}}{M_{\Q}^2}\exp(-\kappa\frac{\langle P_T \rangle^2}{M_{\Q}^2})\left(1+\frac{\kappa}{n}\frac{P_T^2-\langle P_T \rangle^2}{M_{\Q}^2}\right)^{-n}
& \mbox{when $P_T> \langle P_T\rangle$} \\
\end{array}.\nn
\right.\label{eq:crystalball}
}
 $\kappa$,$\lambda$ are fitted via (differential) cross sections obtained from $|\mathcal{A}_{gg\to\Q+X}|^2$
(convoluted with PDFs) to the corresponding experimental data. 
We have used 3 fits:\\
\textbf{Fit 1}: follows from Ref.~\cite{Kom:2011bd}. $\kappa,\lambda$ are obtained through a combined fit of $d\sigma_{\psi}/dP_T$ to the LHCb~\cite{Aaij:2011jh}, ATLAS~\cite{Aad:2011sp}, CMS~\cite{Khachatryan:2010yr} and CDF~\cite{Acosta:2004yw} data.\\
\textbf{Fit 2}: is obtained after including updated CMS measurements~\cite{Chatrchyan:2011kc} covering larger $P_T$.\\
\textbf{Fit 3}: is obtained from a fit of $d^2\sigma_{\psi}/dP_Tdy$  of the CDF~\cite{Acosta:2004yw} data alone.

\begin{table}[hbt!]
\begin{center}\vspace*{-0.25cm}
\begin{tabular}{c|cc}\footnotesize
          & $\kappa$ &$\lambda$ \\
\hline\hline
Fit 1 \cite{Kom:2011bd} & $0.60$  & $0.33$\\
Fit 2 & $0.65$ & $0.32$\\
Fit 3  &  $0.81$ & $0.34$ \\
\end{tabular}\label{cfit}\vspace*{-0.5cm}
\end{center}
\caption{$\kappa$ and $\lambda$ from the 3 fits, with $n=2$ and $\langle P_T\rangle = 4.5$ GeV fixed.}
\end{table}

\vspace*{-0.75cm}
\section{Additional plot}
\label{sec:Ptmin-plot}

\begin{figure}[h!]
\begin{center}
\includegraphics[width=.8\columnwidth,draft=false]{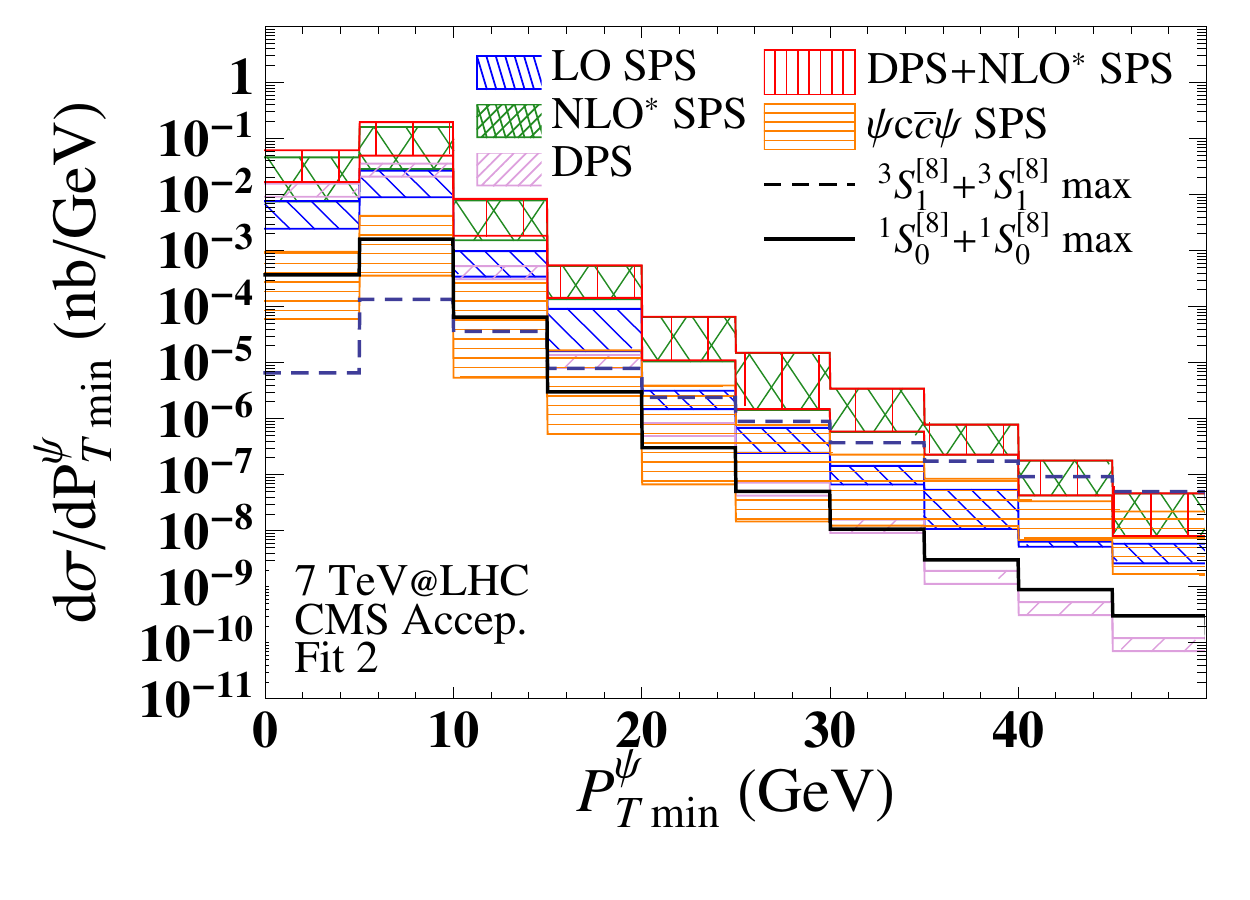}
\caption{Prediction with the CMS kinematical cuts of the $P_{T,\rm min}$ distribution}\label{fig:dsigCMSd}\vspace*{-0.5cm}
\end{center}
\end{figure}


\newpage

\onecolumn
\center{\bf \Large Supplementary materials}

\setcounter{figure}{0}
\setcounter{section}{0}
\setcounter{page}{1}

\vspace*{4cm}

As supplementary materials, we gather  (i) comparisons of the DPS cross section obtained with the Fit 1, 2 \& 3 
and (ii) predictions for forthcoming analyses of data on tapes. We present our predictions  (i) for the ATLAS acceptance in \cf{fig:CompareATLAS}, 
(ii) for the D0 acceptance in ~\cf{fig:CompareDZero}, and (iii) in the expected~\cite{private-LHCb} LHCb rapidity region, 
$2.0<y_{\psi}<4.2$, for the update at $\sqrt{s}=7$ TeV of their earlier measurement~\cite{Aaij:2011yc},
 with a maximum $P_T$ cut, $P_T^{\psi}<10$~GeV in \cf{fig:CompareLHCb}. [We refrain to show $\Delta \phi$ distributions since they are very 
sensitive on the effect of initial-state radiations (or $k_T$ smearing).]

\begin{figure}[hbt!]
\begin{center}
\subfloat[]{\includegraphics[width=.49\columnwidth,draft=false]{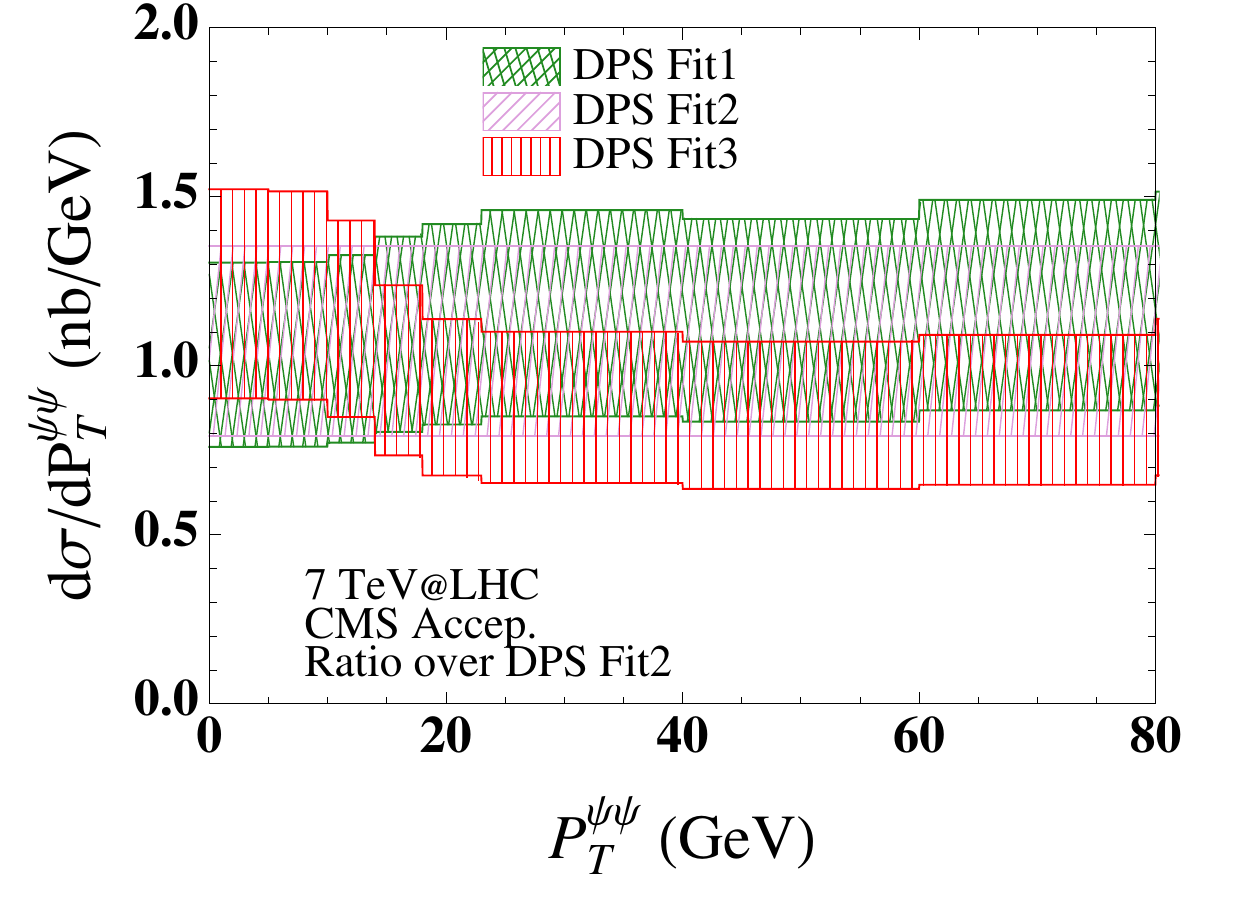}\label{fig:DPSratioa}}\vspace*{-0.3cm}
\subfloat[]{\includegraphics[width=.49\columnwidth,draft=false]{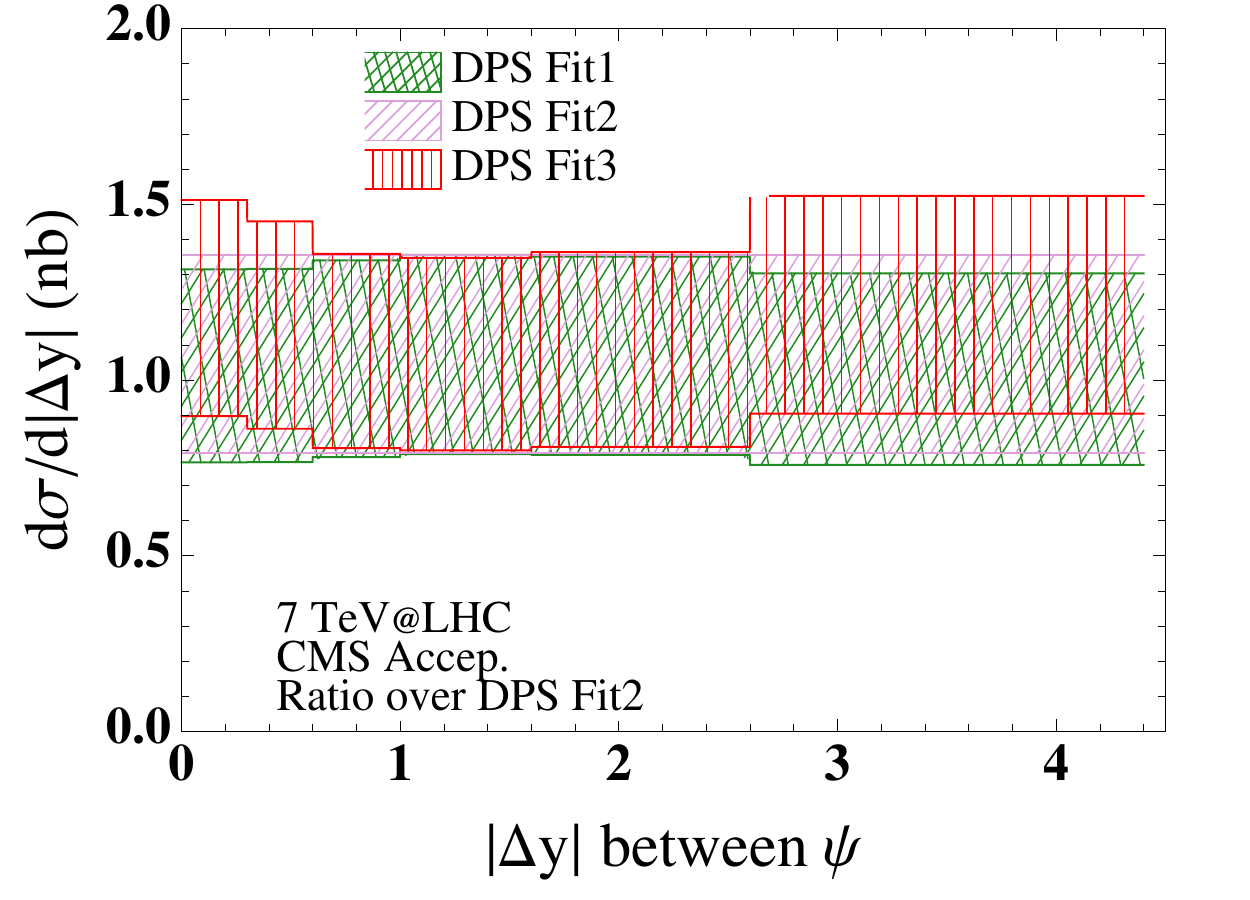}\label{fig:DPSratiob}}\vspace*{-0.3cm}\\
\subfloat[]{\includegraphics[width=.49\columnwidth,draft=false]{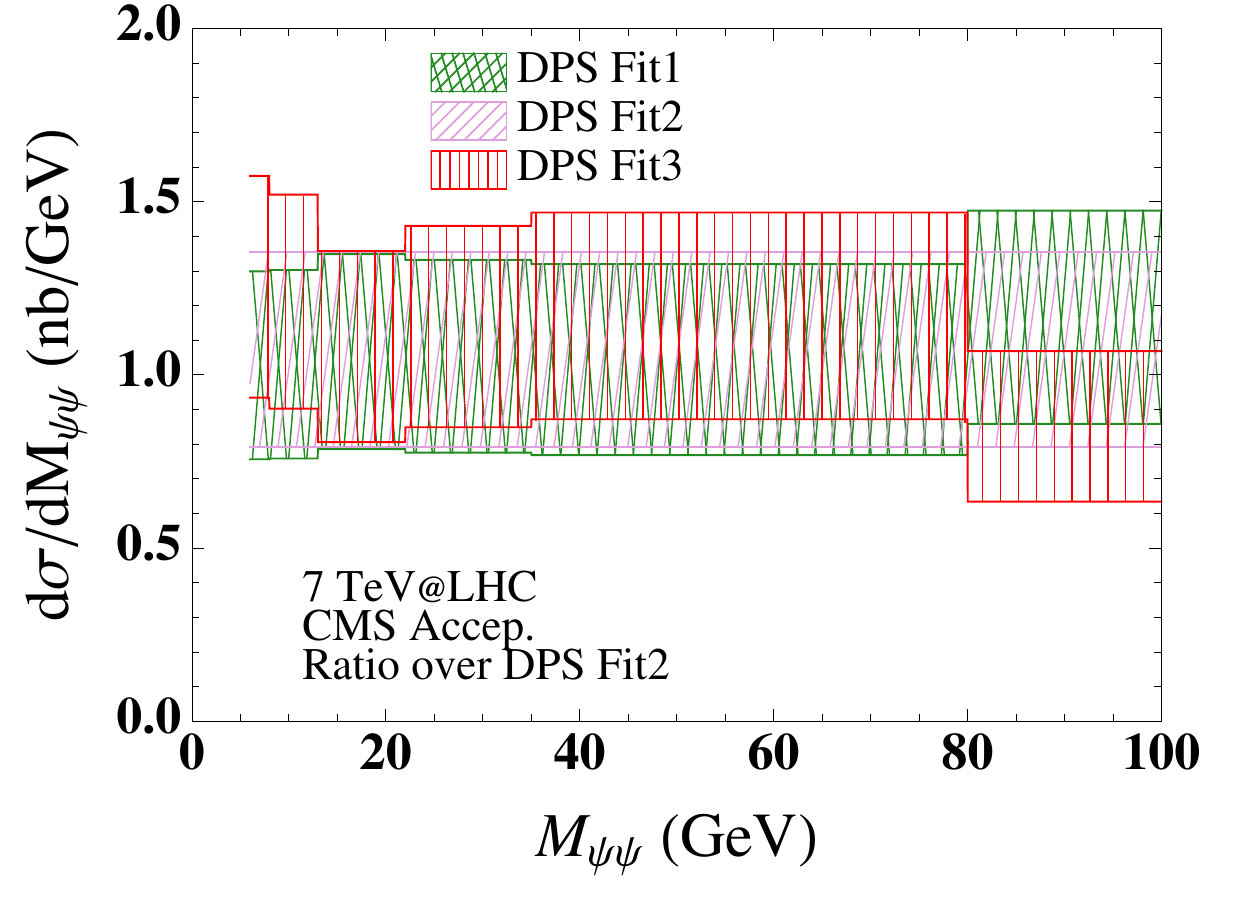}\label{fig:DPSratioc}}\vspace*{-0.3cm}
\
\caption{Ratio of the DPS differential cross section obtained in the CMS acceptance with Fit 1 \& 3 vs Fit 2 as a function of 
(a) transverse momentum spectrum; (b) absolute rapidity difference ; (c) invariant mass distribution.
}
\label{fig:CompareCMSb}
\end{center}\vspace*{-1cm}
\end{figure}

\begin{figure}[hbt!]
\begin{center}
\subfloat[]{\includegraphics[width=.49\columnwidth,draft=false]{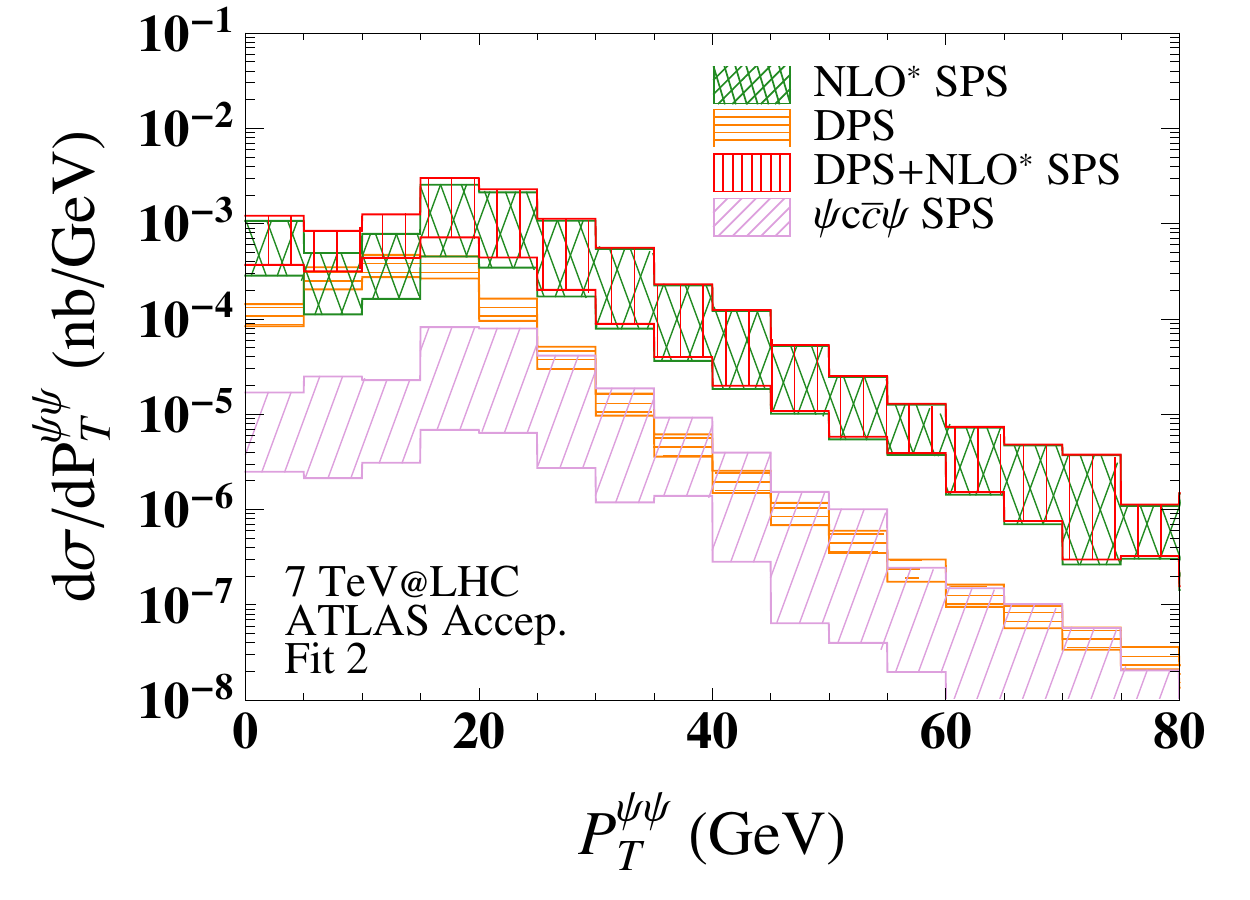}\label{fig:dsigATLASa}}\vspace*{-0.3cm}
\subfloat[]{\includegraphics[width=.49\columnwidth,draft=false]{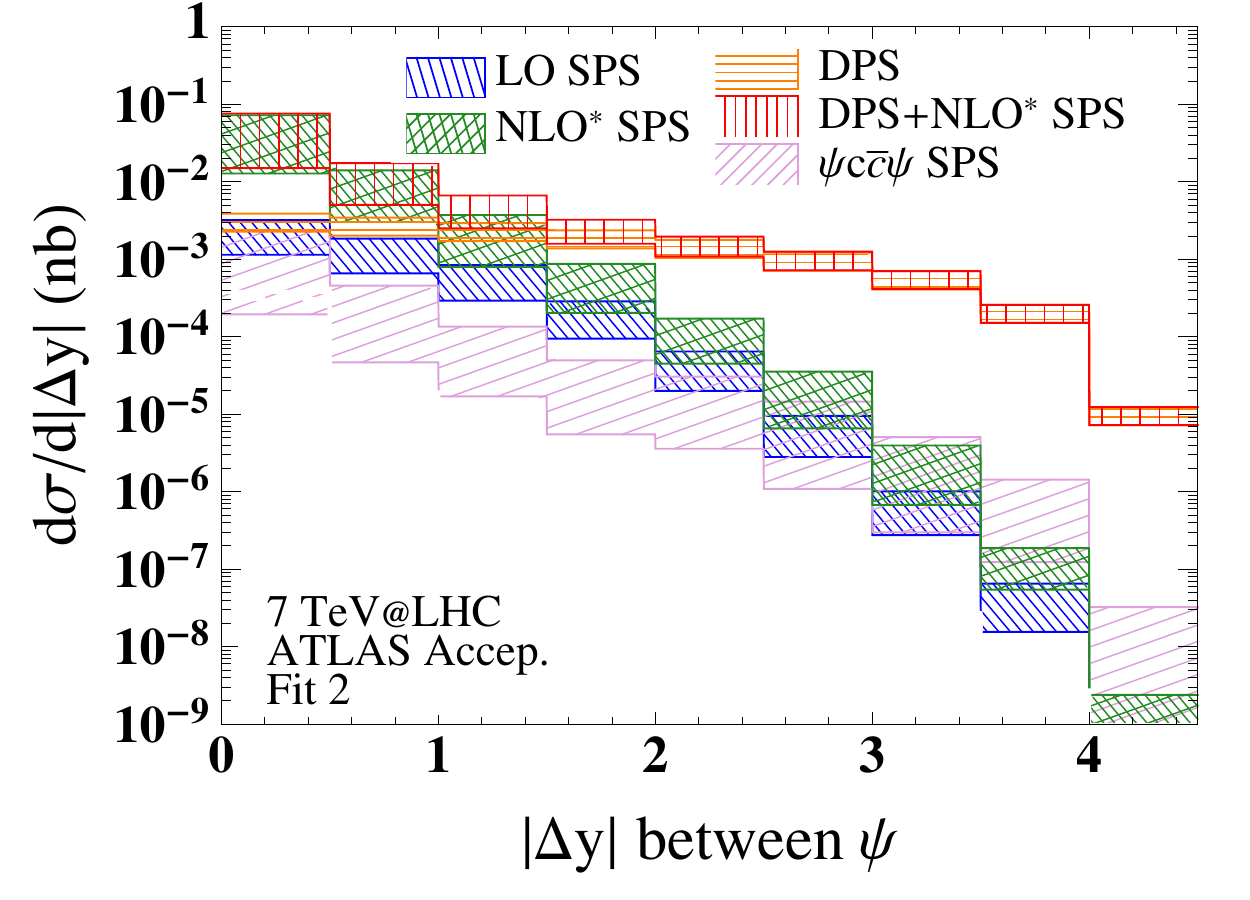}\label{fig:dsigATLASb}}\vspace*{-0.3cm}\\
\subfloat[]{\includegraphics[width=.49\columnwidth,draft=false]{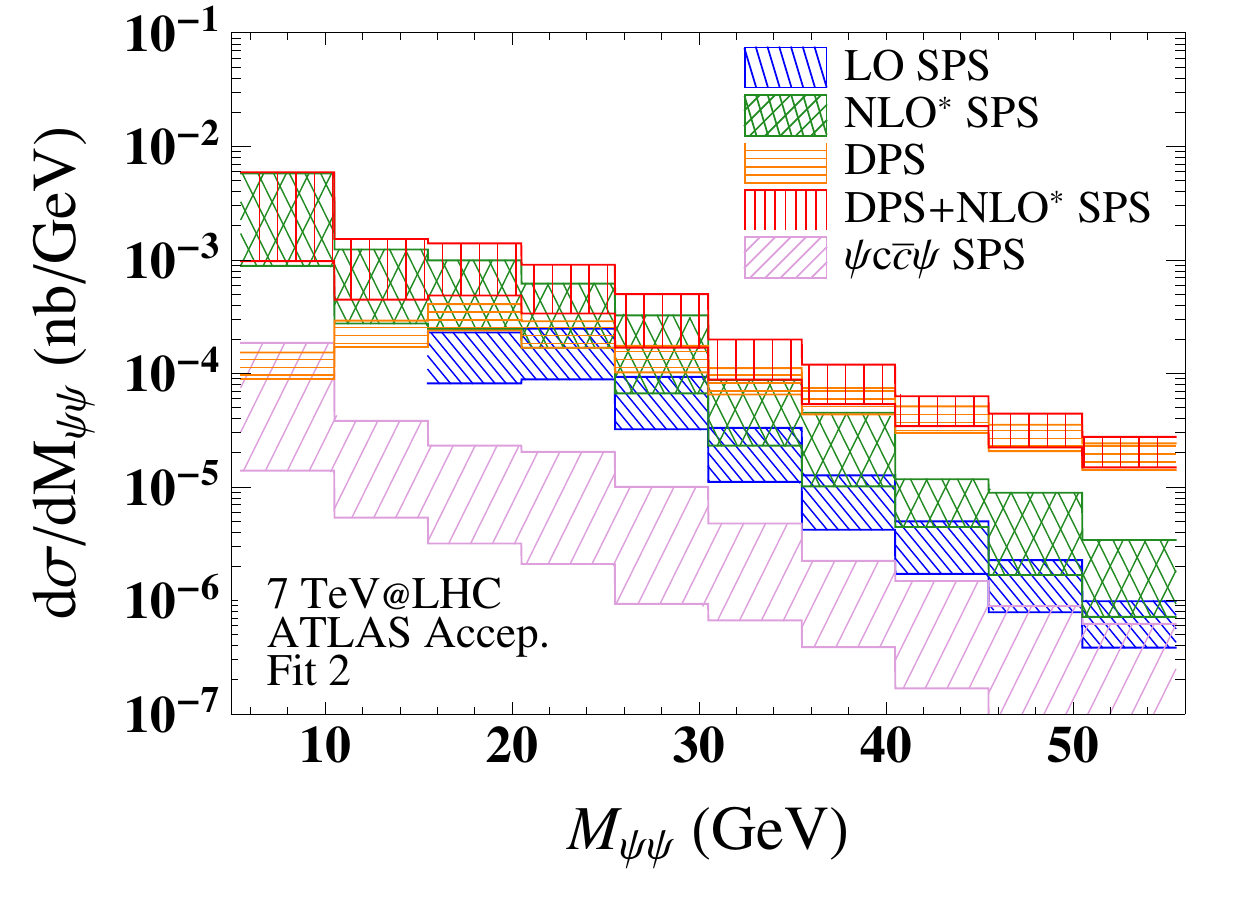}\label{fig:dsigATLASc}}\vspace*{-0.3cm}
\subfloat[]{\includegraphics[width=.49\columnwidth,draft=false]{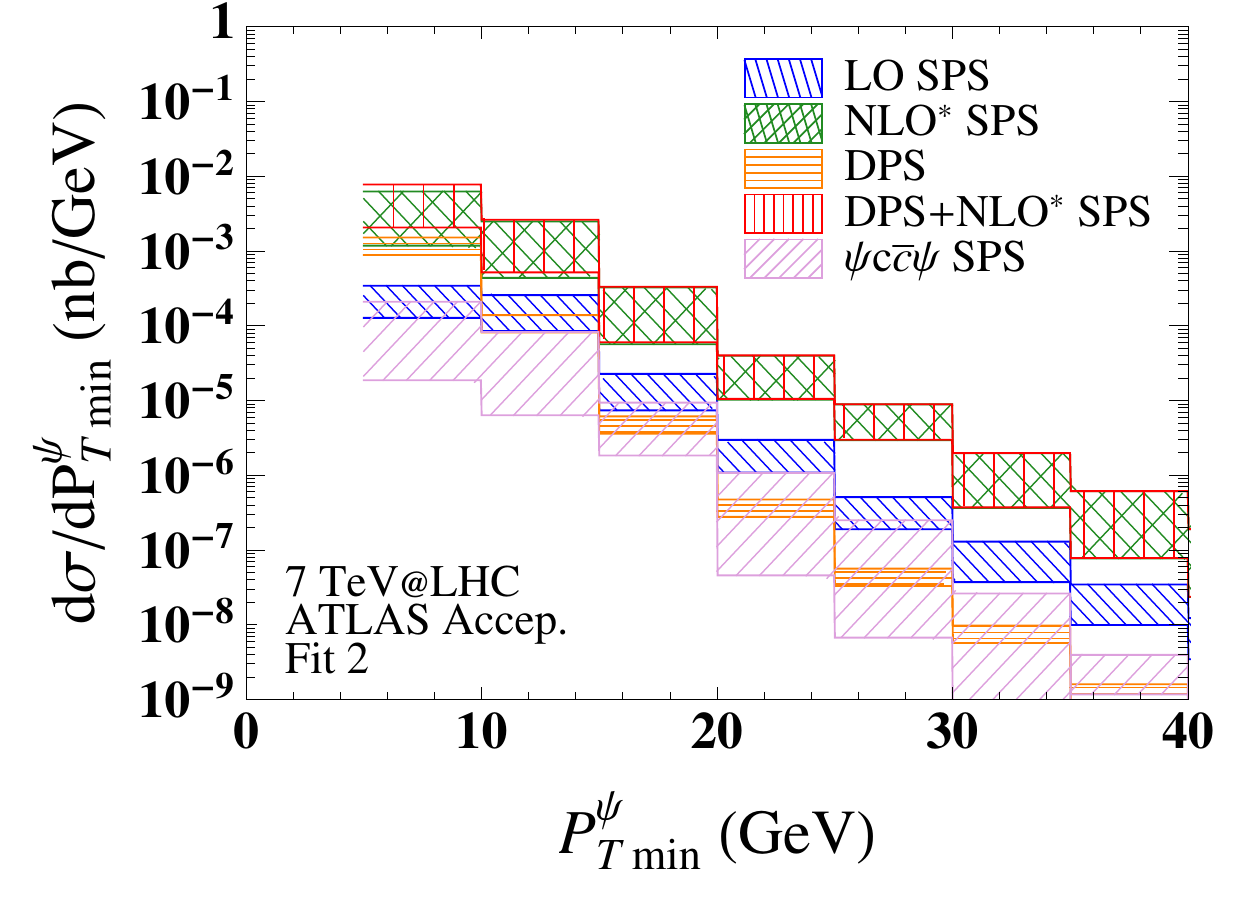}\label{fig:dsigATLASd}}
\caption{Predictions with the ATLAS kinematical cuts:(a) transverse momentum spectrum; (b) absolute rapidity difference ; (c) invariant mass distribution;
(d) sub-leading $P_T$ spectrum.
}
\label{fig:CompareATLAS}
\end{center}\vspace*{-1cm}
\end{figure}

\begin{figure}[hbt!]
\begin{center}
\subfloat[]{\includegraphics[width=.49\columnwidth,draft=false]{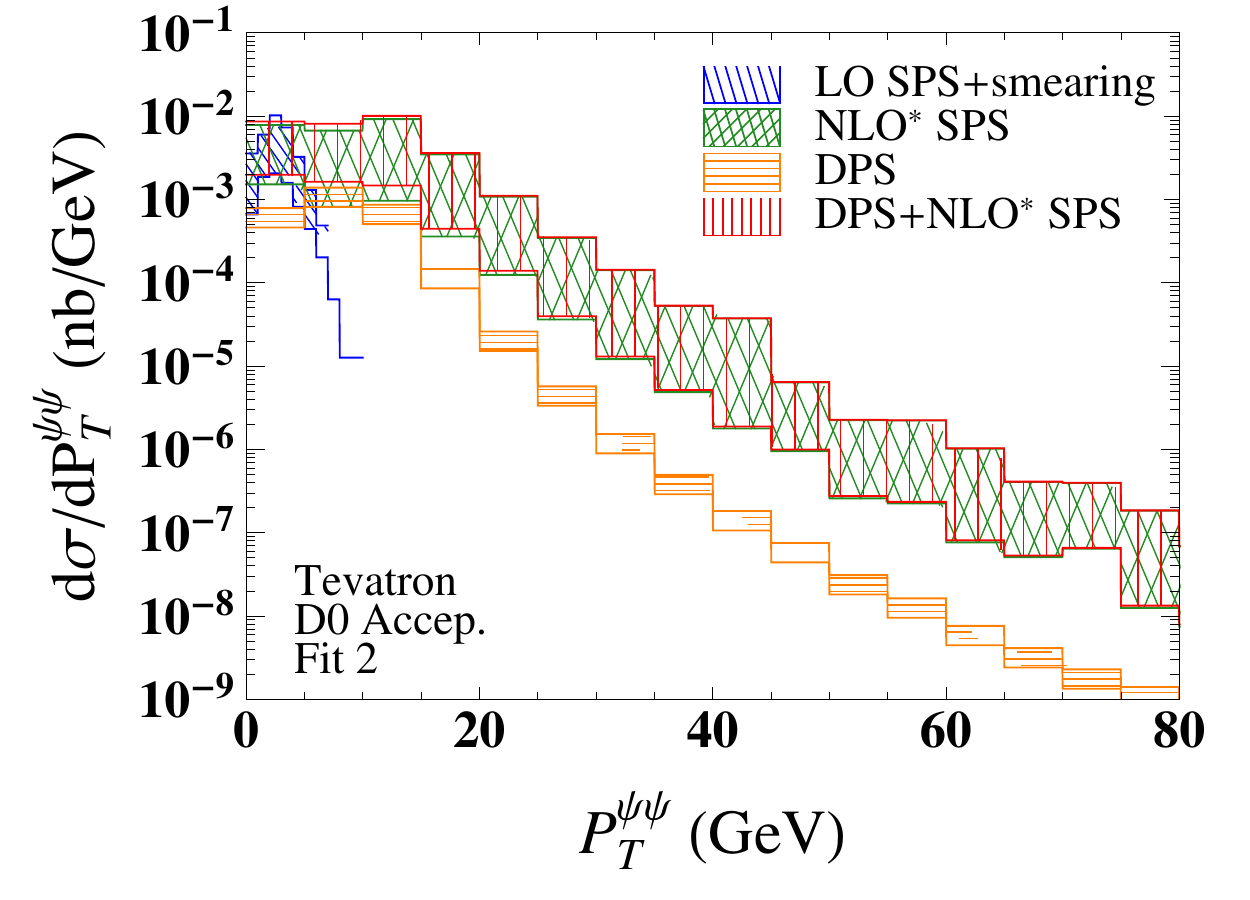}\label{fig:dsigDZeroa}}\vspace*{-0.3cm}
\subfloat[]{\includegraphics[width=.49\columnwidth,draft=false]{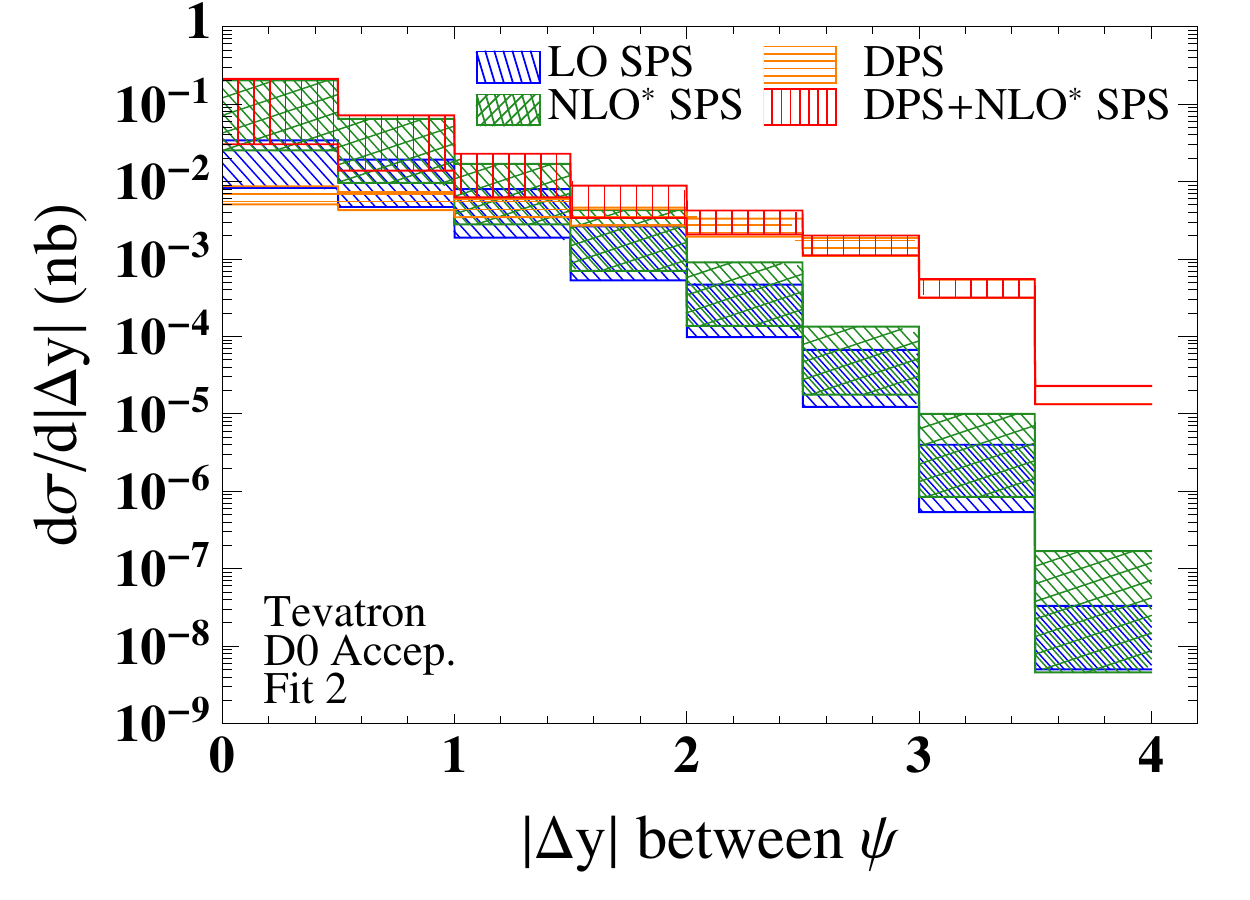}\label{fig:dsigDZerob}}\vspace*{-0.3cm}\\
\subfloat[]{\includegraphics[width=.49\columnwidth,draft=false]{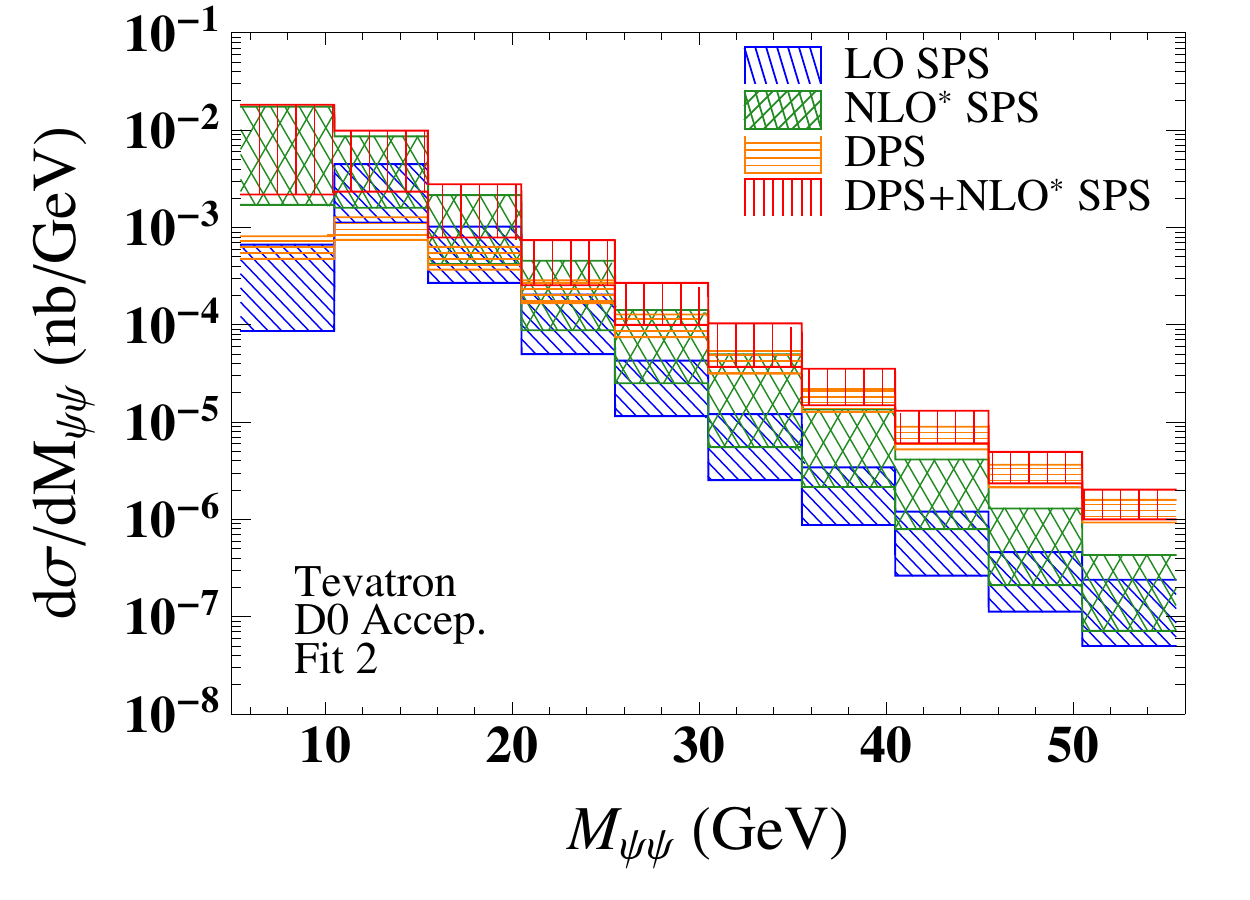}\label{fig:dsigDZeroc}}\vspace*{-0.3cm}
\subfloat[]{\includegraphics[width=.49\columnwidth,draft=false]{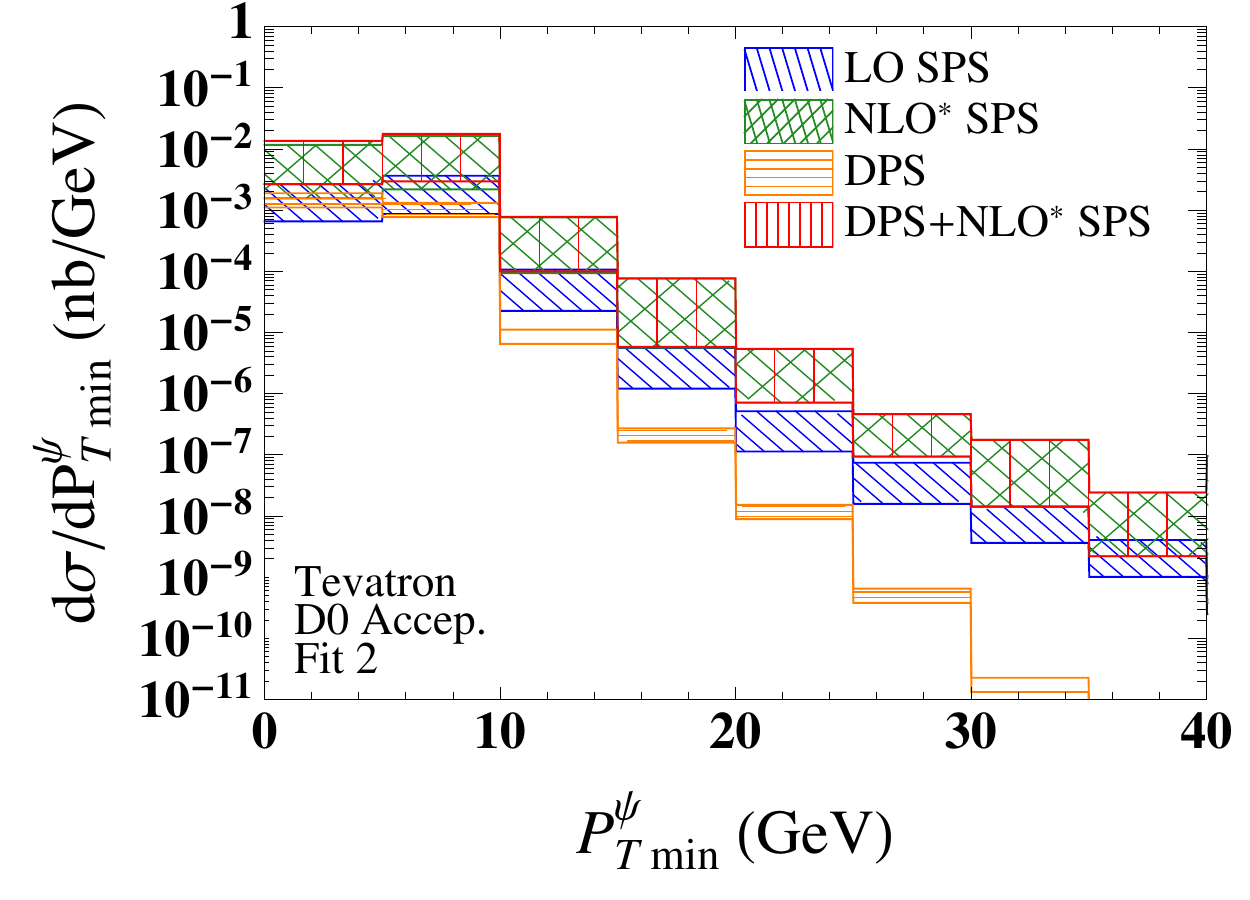}\label{fig:dsigDZerod}}
\caption{Predictions with the D0 kinematical cuts:(a) transverse momentum spectrum; (b) absolute rapidity difference ; (c) invariant mass distribution;
(d) sub-leading $P_T$ spectrum.}
\label{fig:CompareDZero}
\end{center}\vspace*{-1cm}
\end{figure}

\begin{figure}
\begin{center}
\subfloat[]{\includegraphics[width=.49\columnwidth,draft=false]{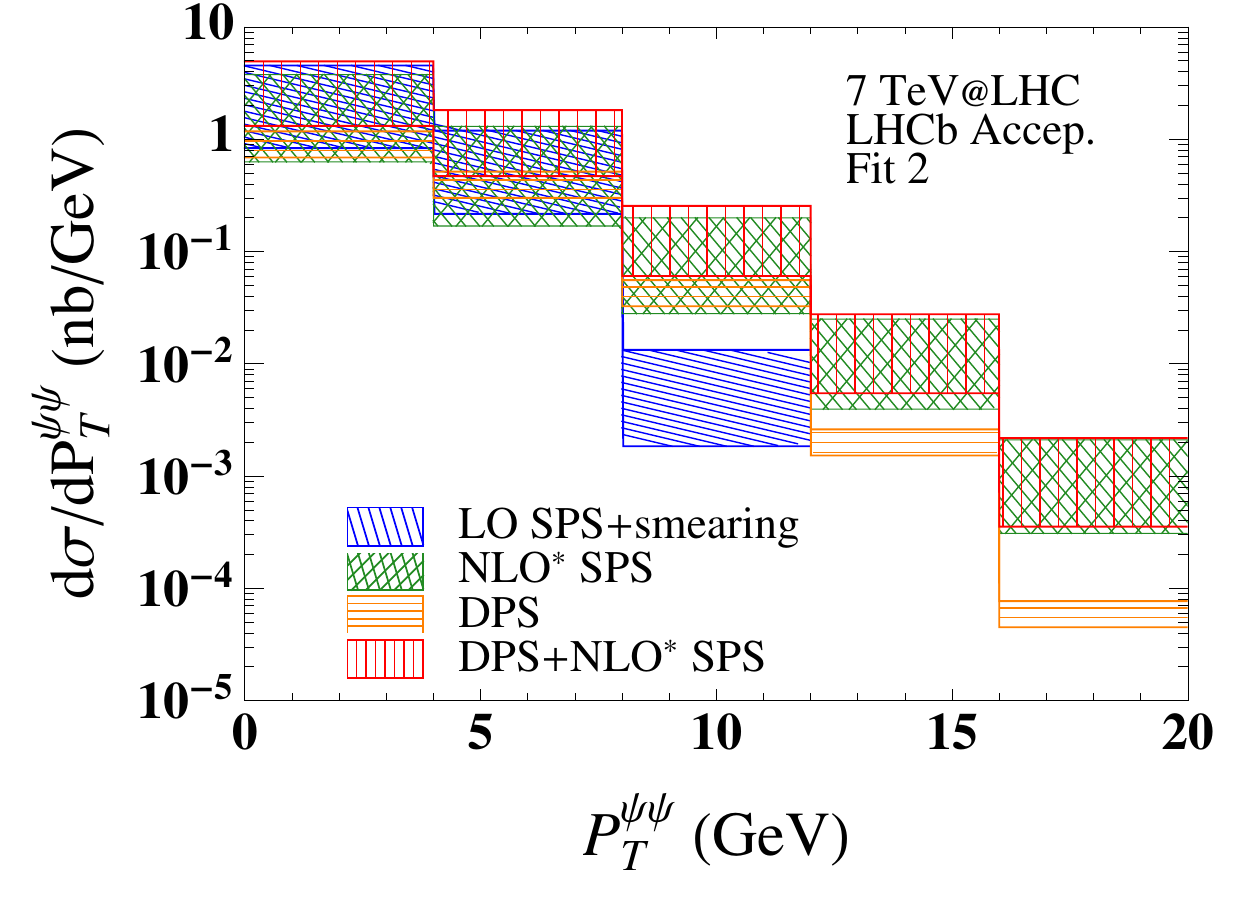}\label{fig:dsigLHCba}}\vspace*{-0.3cm}
\subfloat[]{\includegraphics[width=.49\columnwidth,draft=false]{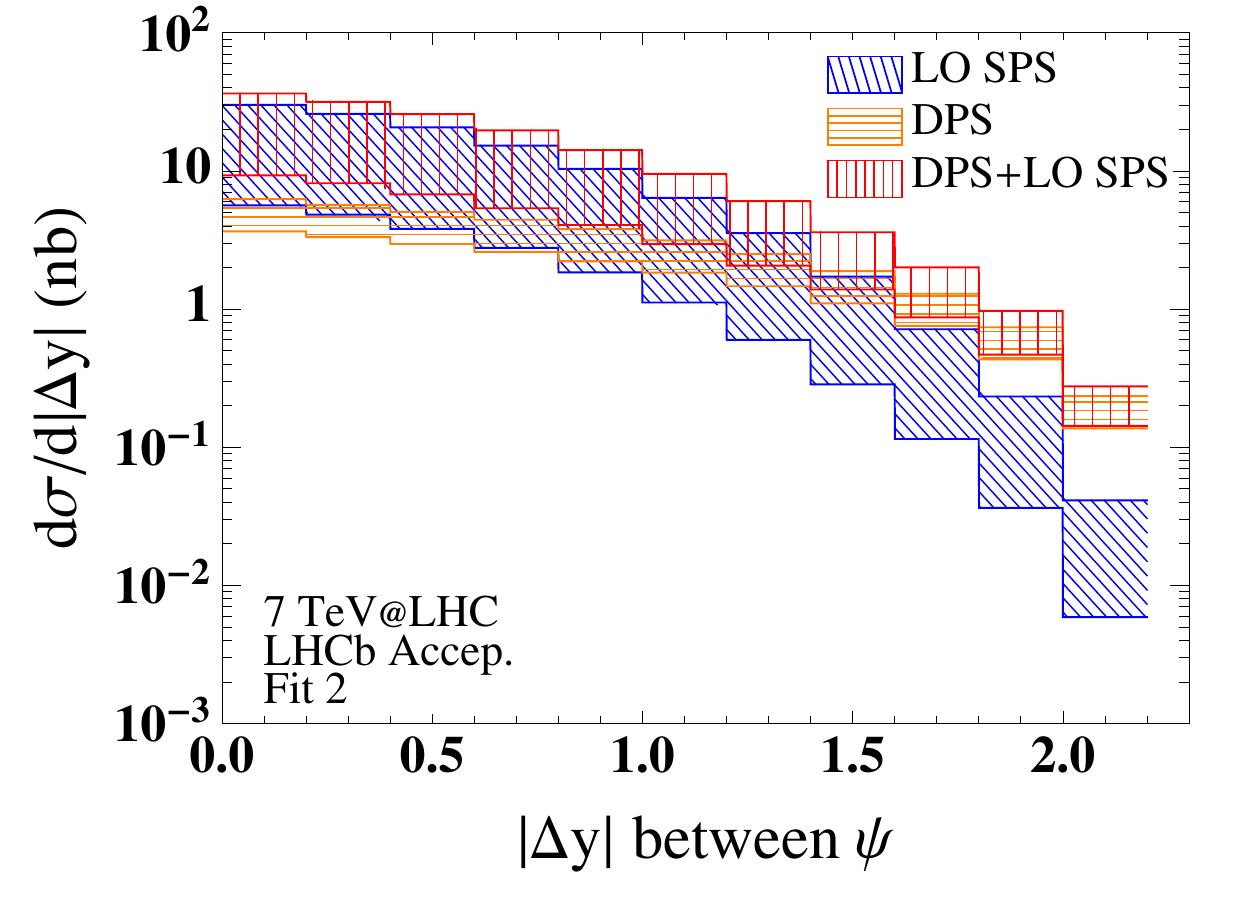}\label{fig:dsigLHCbb}}\vspace*{-0.3cm}\\
\subfloat[]{\includegraphics[width=.49\columnwidth,draft=false]{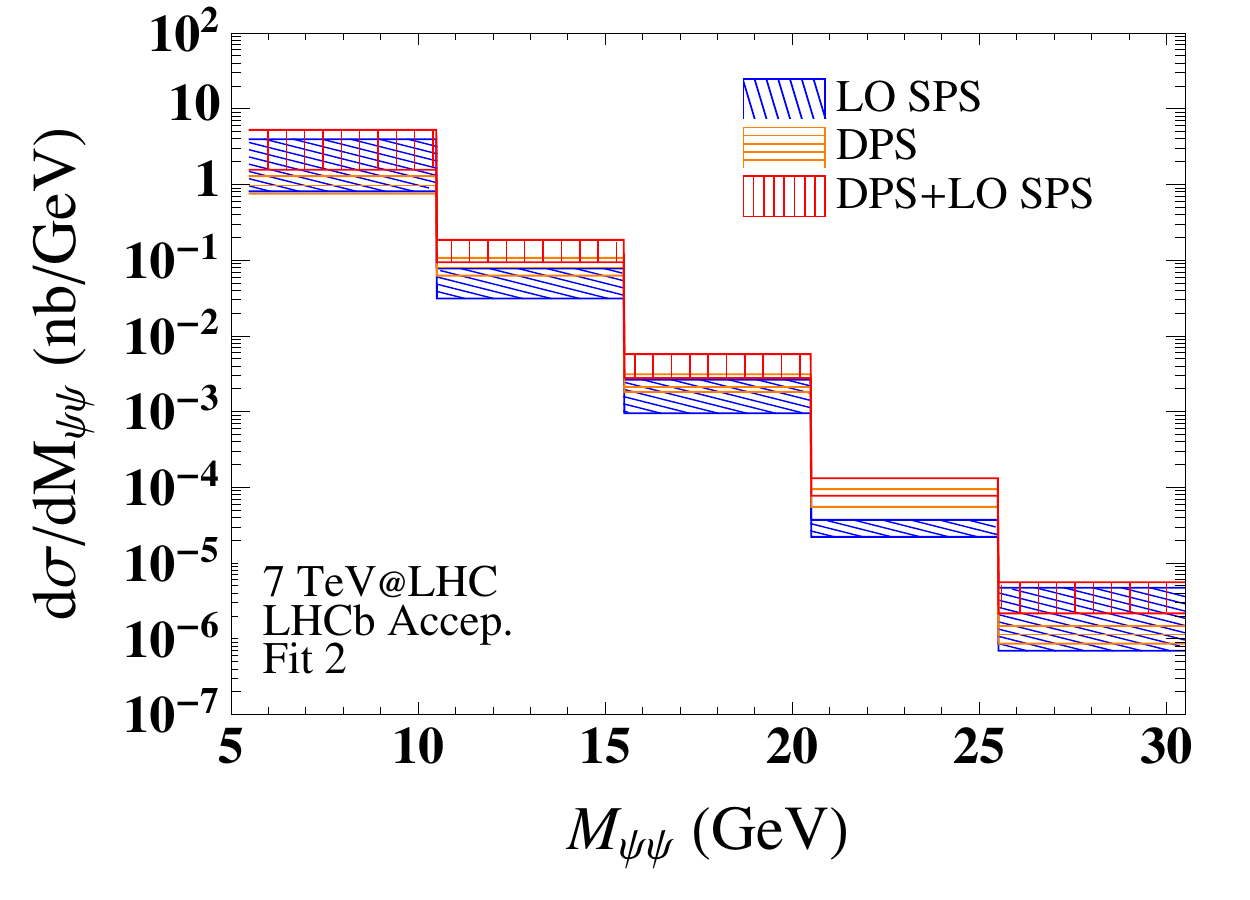}\label{fig:dsigDLHCbc}}\vspace*{-0.3cm}
\subfloat[]{\includegraphics[width=.49\columnwidth,draft=false]{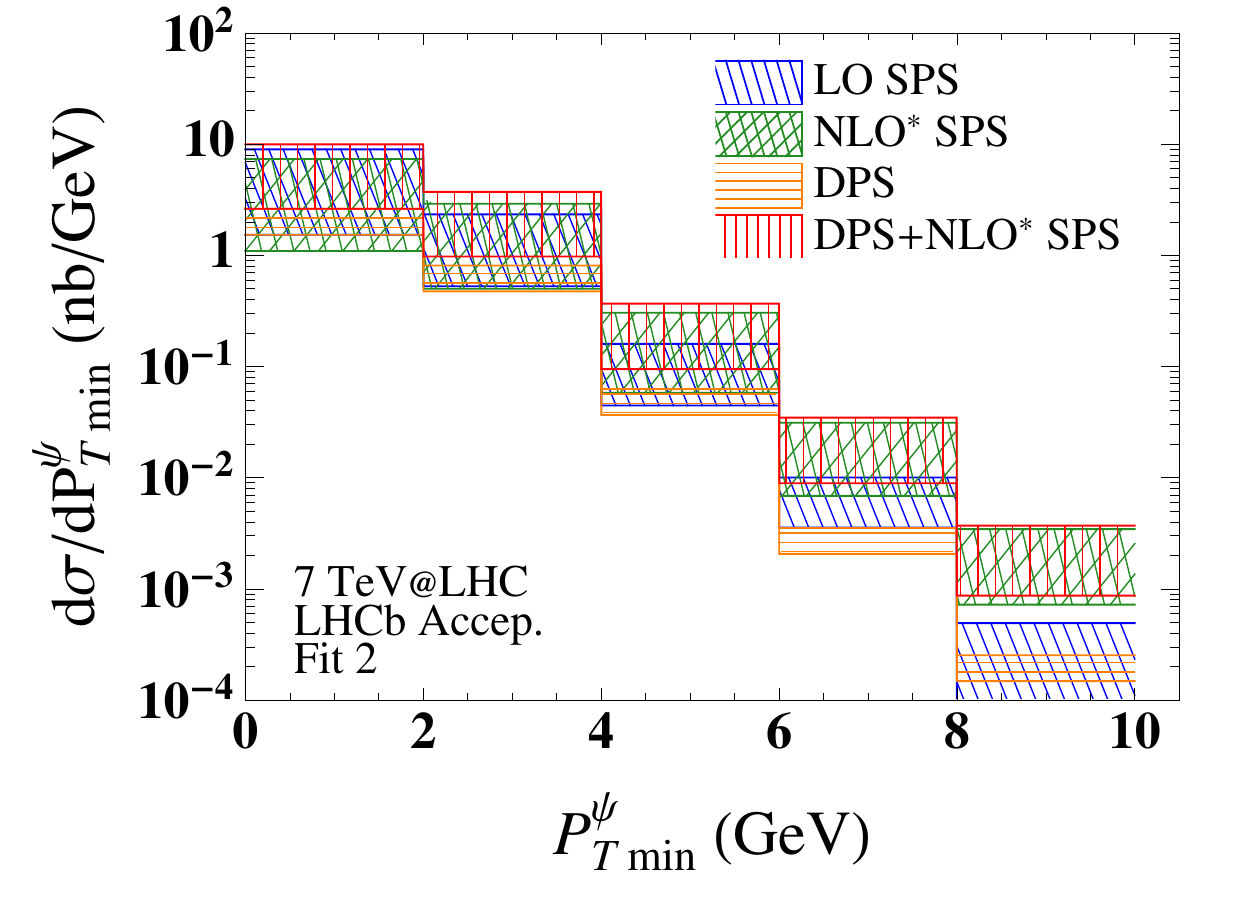}\label{fig:dsigLHCbd}}
\caption{Predictions with the LHCb kinematical cuts:(a) transverse momentum spectrum; (b) absolute rapidity difference ; (c) invariant mass distribution;
(d) sub-leading $P_T$ spectrum.}
\label{fig:CompareLHCb}
\end{center}\vspace*{-1cm}
\end{figure}

\end{document}